\documentclass[10pt]{article}
\usepackage{amssymb}
\newcommand{\nc}{\newcommand}
\nc{\bib}{\bibitem}
\nc{\al}{\alpha}
\nc{\g}{\gamma}
\nc{\G}{\Gamma}
\nc{\D}{\Delta}
\nc{\Ups}{\Upsilon}
\nc{\eps}{\epsilon}
\nc{\la}{\lambda}
\nc{\La}{\Lambda}
\nc{\var}{\varphi}
\nc{\thh}{\hat{\theta}}
\nc{\Dh}{\hat{\Delta}}
\nc{\Ah}{\hat{A}}
\nc{\Bh}{\hat{B}}
\nc{\Dha}{\hat{D}}
\nc{\pa}{\partial}
\nc{\nn}{\nonumber \\ }
\nc{\hf}{\frac{1}{2}}  
\nc{\dz}{\frac{dz}{2\pi i}}
\nc{\bin}[2]{\left (\begin{array}{c} {#1}\\ {#2} \end{array}\right )}
\nc{\be}{\begin{equation}}
\nc{\ee}{\end{equation}}
\nc{\bea}{\begin{eqnarray}}
\nc{\eea}{\end{eqnarray}}
\nc{\bra}[1]{\langle {#1}|}
\nc{\ket}[1]{|{#1}\rangle}

\def\vvdots{\mathinner{\mkern1mu\raise1pt\vbox{\kern7pt\hbox{.}}\mkern2mu
 \raise4pt\hbox{.}\mkern2mu\raise7pt\hbox{.}\mkern1mu}}

\begin{document}

\topmargin -5mm
\oddsidemargin 5mm

\begin{titlepage}
\setcounter{page}{0}

\vspace{8mm}
\begin{center}
{\Huge Affine Jordan cells, logarithmic correlators,}\\[.4cm]
{\Huge and hamiltonian reduction}

\vspace{15mm}
{\huge J{\o}rgen Rasmussen}\\[.3cm] 
{\large {\em Department of Mathematics and Statistics, University of Concordia}}\\ 
{\large {\em 1455 Maisonneuve W, Montr\'eal, Qu\'ebec, Canada H3G 1M8}}\\[.3cm]
{\large rasmusse@crm.umontreal.ca}

\end{center}

\vspace{10mm}
{\large
\centerline{{\bf{Abstract}}}    }
\vskip.4cm
\noindent
We study a particular type of logarithmic extension of $SL(2,\mathbb{R})$
Wess-Zumino-Witten models. It is based on the introduction of affine Jordan cells
constructed as multiplets of quasi-primary fields organized in 
indecomposable representations of the Lie algebra $sl(2)$.
We solve the simultaneously imposed set of
conformal and $SL(2,\mathbb{R})$ Ward identities
for two- and three-point chiral blocks. These correlators will in general involve
logarithmic terms and may be represented compactly by considering
spins with nilpotent parts. The chiral blocks are found to
exhibit hierarchical structures revealed by computing derivatives
with respect to the spins.
We modify the Knizhnik-Zamolodchikov equations to cover affine Jordan cells
and show that our chiral blocks satisfy these equations.
It is also demonstrated that
a simple and well-established prescription for hamiltonian reduction at the level
of ordinary correlators extends straightforwardly to the logarithmic 
correlators as the latter then reduce to the known results for
two- and three-point conformal blocks in logarithmic conformal field theory.
\\[.5cm]
{\bf Keywords:} Logarithmic conformal field theory, Jordan cell,
Wess-Zumino-Witten model, Knizhnik-Zamolodchikov equations,
hamiltonian reduction.
\end{titlepage}
\newpage
\renewcommand{\thefootnote}{\arabic{footnote}}
\setcounter{footnote}{0}

\section{Introduction}

In logarithmic conformal field theory (CFT), a primary field may
have a so-called
logarithmic partner field on which the Virasoro modes do not all act
diagonally. If only one logarithmic field is associated to a given
primary field, the two fields constitute a 
so-called conformal Jordan cell of rank two where the rank indicates the number
of fields in the cell. We will be concerned with 
conformal Jordan cells of rank two only. The appearance of such cells
is known to lead to logarithmic singularities in the correlators.
We refer to \cite{Gur}
for the first systematic study of logarithmic CFT, and to
\cite{Flo,Gab,Nic} for recent reviews on the subject. 
An exposition of links to string theory may be found in \cite{Mav}.

The objective of the present work is to introduce and study a
particular logarithmic extension of the $SL(2,\mathbb{R})$
Wess-Zumino-Witten (WZW) model.
Alternative extensions have appeared in the literature,
see \cite{Gabsu2,FFHST,LMRS,Nic,Nicsu2}, for example,
but all seem to differ significantly from ours in foundation
and approach. 

Our construction is based on a generalization
of the standard multiplets of Virasoro primary fields organized
as spin-$j$ representation. 
We find that an infinite number of partner fields
seem to be required to complete such an indecomposable
representation of $sl(2)$, and we refer to these new
multiplets as affine Jordan cells.

We consider the case where the logarithmic fields
in the affine Jordan cells are quasi-primary, and
discuss the conformal and $SL(2,\mathbb{R})$
Ward identities which follow.
Without making any simplifying assumptions about the 
operator-product expansions of the fields, we find the general
solutions for two- and three-point chiral blocks. Our results
thus cover all the possible cases based on 
primary fields not belonging to affine
Jordan cells, primary fields belonging to affine Jordan
cells, and the logarithmic partner fields completing
the affine Jordan cells. 

Most of our computations are based on the introduction
of generating functions for the fields appearing in the
various representations. This means that the 
affine correlators of the individual fields
are obtained by expanding certain
generating-function chiral blocks. 

A modification of the Knizhnik-Zamolodchikov (KZ) equations 
\cite{KZ}
is required to cover affine Jordan cells in addition to primary fields.
It is demonstrated that the chiral blocks obtained as solutions
to the Ward identities 
satisfy these generalized KZ equations.
This verification is straightforward once it has
been established that the two- and three-point
chiral blocks may be expressed compactly
in terms of spins with nilpotent parts. We show
that this is possible.
It also follows that a two- or three-point
chiral block factorizes into a `conformal` part
and a `group` part.

The chiral blocks are found to
exhibit hierarchical structures obtained by computing derivatives
with respect to the spins.
This extends an observation made in \cite{RAK,FloOPE,logWard}
that the sets of two- and three-point conformal blocks
in logarithmic CFT are linked via derivatives with respect to
the conformal weights.

It is noted that a simple distinction has been introduced as we refer to
chiral correlators in the WZW model
as {\em chiral blocks}, while chiral correlators
in logarithmic CFT are referred to as {\em conformal blocks}.
This is common practice.

A merit of our construction seems to be that the 
affine correlators reduce to the conformal ones when a
straightforward extension of the 
prescription for hamiltonian reduction introduced in
\cite{FGPP,GP} is employed. The idea is formulated in the realm
of generating functions for the Virasoro primary fields
in spin-$j$ multiplets, and we find that it may be
extended to affine Jordan cells and thus cover
the reduction of our logarithmic $SL(2,\mathbb{R})$
WZW model to logarithmic CFT.

This paper proceeds as follows. 
To fix our notation and to prepare for the discussion of hamiltonian
reduction, we first review the recently obtained general
solutions to the conformal Ward identities for two- and
three-point conformal blocks in logarithmic
CFT \cite{logWard}. 
This is followed by a discussion in Section 3 of generating-function
primary fields and their correlators in $SL(2,\mathbb{R})$ WZW models.
This is the framework which we extend in Section \ref{affJc} and eventually
use in our analysis of chiral blocks.
In Section \ref{affJc}, we thus describe the indecomposable $sl(2)$
representations underlying the affine Jordan cells.
The correspondingly modified KZ equations are also
introduced. Section \ref{corr} concerns the explicit results on
two- and three-point chiral blocks. It includes a discussion
of the factorization of the chiral blocks based on
spins with nilpotent parts,
as well as a discussion of the hierarchical structures of the chiral blocks.
Some technical details are deferred to Appendix \ref{app}.
The extended prescription for hamiltonian reduction is considered
in Section \ref{hamred}, where it is demonstrated that our
chiral blocks reduce to the conformal ones reviewed in Section 
\ref{LogCFT}.
Section \ref{concl} contains some concluding remarks.

\section{Correlators in logarithmic CFT}
\label{LogCFT}

\subsection{Conformal Jordan cell}

A conformal Jordan cell of rank two consists of two fields:
a primary field, $\Phi$, of conformal weight $\Dh$ and
its non-primary, `logarithmic' partner field, $\Psi$, 
on which the Virasoro algebra 
\be
 [L_n,L_m]=(n-m)L_{n+m}+\frac{\hat{c}}{12}n(n^2-1)\delta_{n+m,0}
\label{vir}
\ee
generated by $\{L_n\}$ does not act diagonally. The central extension
is denoted $\hat{c}$.
With a conventional relative normalization of the fields, we have
\bea
 \left[L_n,\Phi(z)\right]&=&\left(z^{n+1}\pa_z+\Dh(n+1)z^n\right)\Phi(z)\nn
 \left[L_n,\Psi(z)\right]&=&\left(z^{n+1}\pa_z+\Dh(n+1)z^n\right)\Psi(z)
  +(n+1)z^n\Phi(z)
\label{L}
\eea

It has been suggested by Flohr \cite{Flo-9707} to describe these fields
in a unified way by introducing a nilpotent, yet even, parameter
$\thh$ satisfying $\thh^2=0$. We will follow this idea here,
though use an approach closer to the one employed in \cite{MRS,logsle,logWard}.
We thus define the field or unified cell
\be
 \Ups(z;\thh)\ =\ \Phi(z)+\thh\Psi(z)
\label{Ups}
\ee
which is seen to be `primary' of conformal weight $\Dh+\thh$ as
the commutators (\ref{L}) may be expressed as
\be
 \left[L_n,\Ups(z;\thh)\right]\ =\ 
  \left(z^{n+1}\pa_z+(\Dh+\thh)(n+1)z^n\right)\Ups(z;\thh)
\label{Lup}
\ee

Following \cite{logWard},
a primary field belonging to a conformal Jordan cell is referred to
as a cellular primary field. 
A primary field {\em not} belonging to a conformal Jordan cell may 
be represented as $\Ups(z;0)$, and we will reserve this notation
for these non-cellular primary fields. To avoid ambiguities, we will therefore
refrain from considering unified cells
$\Ups(z;\thh)$, as defined in (\ref{Ups}), for vanishing $\thh$.

\subsection{Conformal Ward identities}

We consider {\em quasi-primary fields} only, ensuring the projective invariance
of their correlators constructed by sandwiching the fields 
between projectively invariant vacua. 
This invariance is made manifest qua 
the conformal Ward identities which are 
given here for ${\cal N}$-point conformal blocks:
\bea
 0&=&\sum_{i=1}^{\cal N}\pa_{z_i}\langle\Ups_1(z_1;\thh_1)\dots
  \Ups_{\cal N}(z_{\cal N};\thh_{\cal N})\rangle\nn
 0&=&\sum_{i=1}^{\cal N}\left(z_i\pa_{z_i}+\Dh_i+\thh_i\right)
  \langle\Ups_1(z_1;\thh_1)\dots
   \Ups_{\cal N}(z_{\cal N};\thh_{\cal N})\rangle\nn
 0&=&\left({\cal L}^{\cal N}+2\sum_{i=1}^{\cal N}\thh_iz_i\right)
   \langle\Ups_1(z_1;\thh_1)\dots
   \Ups_{\cal N}(z_{\cal N};\thh_{\cal N})\rangle
\label{confward}
\eea
To simplify the notation, we have introduced the differential
operator
\be
 {\cal L}^{\cal N}\ =\ \sum_{i=1}^{\cal N}
  \left(z_i^2\pa_{z_i}+2\Dh_iz_i\right)
\label{calL}
\ee

Information on the individual correlators may be
extracted from solutions to the
conformal Ward identities involving unified cells. In the case of
\be
 \langle\Ups_1(z_1;\thh_1)\Ups_2(z_2;0)
  \Ups_3(z_3;\thh_3)\rangle
\ee 
for example, the third conformal Ward identity (\ref{confward}) reads
\be
 0\ =\ \left({\cal L}^3+2(\thh_1z_1+\thh_3z_3)\right)
   \langle\Ups_1(z_1;\thh_1)\Ups_2(z_2;0)
   \Ups_3(z_3;\thh_3)\rangle
\ee
A solution to the complete set of conformal Ward identities 
is an expression expandable in $\thh_1$
and $\thh_3$. The term proportional to $\thh_1$
but independent of $\thh_3$, for example, 
should then be identified
with $\langle\Psi_1(z_1)\Ups_2(z_2;0)\Phi_3(z_3)\rangle$.

By construction, and as illustrated by this example,
correlators involving unified cells and
non-cellular primary fields may thus
be regarded as generating-function correlators whose expansions
in the nilpotent parameters give the individual correlators
involving combinations of cellular primary fields, non-cellular primary 
fields, and logarithmic fields.
Our focus will therefore be on correlators of combinations
of unified cells and non-cellular primary fields.

\subsection{Two-point conformal blocks}

Based on the ansatz
\be
 \langle\Ups_1(z_1;\thh_1)\Ups_2(z_2;\thh_2)\rangle\ 
  =\ \frac{\Ah(\thh_1,\thh_2)+\Bh(\thh_1,\thh_2)\ln z_{12}}{
  z_{12}^{2h}}
\label{2ans}
\ee
where
\be
 \Ah(\thh_1,\thh_2)\ =\ 
  \Ah^0+\Ah^1\thh_1+\Ah^2\thh_2+\Ah^{12}\thh_1\thh_2
\label{Aexp}
\ee
and similarly for $\Bh(\thh_1,\thh_2)$,
the general (generating-function) two-point conformal blocks read
\bea
   \langle\Ups_1(z_1;0)\Ups_2(z_2;0)\rangle
 &=&\Ah^0V_2\nn
  \langle\Ups_1(z_1;\thh_1)\Ups_2(z_2;0)\rangle
 &=&\Ah^1\thh_1V_2\nn
  \langle\Ups_1(z_1;0)\Ups_2(z_2;\thh_2)\rangle
  &=&\Ah^2\thh_2V_2\nn
  \langle\Ups_1(z_1;\thh_1)\Ups_2(z_2;\thh_2)\rangle
 &=&\left\{\Ah^1\thh_1+\Ah^1\thh_2+\left(\Ah^{12}
  -2\Ah^1\ln z_{12}\right)\thh_1\thh_2\right\}V_2
\label{2uni}
\eea
Here we have introduced the shorthand notation
\be
 V_2\ =\ \frac{\delta_{\Dh_1,\Dh_2}}{z_{12}^{\Dh_1+\Dh_2}}
\label{V2}
\ee
To keep the notation simple, we are using the standard
abbreviation $z_{ij}=z_i-z_j$.
It is understood that an $\Ah^1$, for example, appearing in 
one (generating-function) correlator
a priori is independent of an $\Ah^1$ appearing in another.
Also, even though $\Ah^2$ does not appear explicitly
in some of these expressions, it may nevertheless be related to $\Ah^1$.
For the sake of simplicity, the solutions listed here are merely
indicating the general form and the degrees of freedom
without reference to the fate of all the various parameters  
appearing in the ansatz (\ref{2ans}). Similar comments also apply
to the results on correlators discussed in the following.
Finally, the solutions for the individual two-point conformal blocks are easily
extracted \cite{logWard}.

By considering $\thh_i$ as the nilpotent part of the generalized
conformal weight $\Dh_i+\thh_i$ \cite{MRS,logWard}, 
one may represent the results (\ref{2uni}) as
\bea
  \langle\Ups_1(z_1;0)\Ups_2(z_2;0)\rangle
 &=&\delta_{\Dh_1,\Dh_2}\frac{\Ah^0}{z_{12}^{\Dh_1+\Dh_2}}\nn
   \langle\Ups_1(z_1;\thh_1)\Ups_2(z_2;0)\rangle
 &=&\delta_{\Dh_1,\Dh_2}\frac{\Ah^1\thh_1}{z_{12}^{(\Dh_1+\thh_1)+\Dh_2}}\nn
   \langle\Ups_1(z_1;\thh_1)\Ups_2(z_2;\thh_2)\rangle
 &=&\delta_{\Dh_1,\Dh_2}\frac{\Ah^1\thh_1+\Ah^1\thh_2
  +\Ah^{12}\thh_1\thh_2}{z_{12}^{(\Dh_1+\thh_1)+(\Dh_2+\thh_2)}}
\label{2unith}
\eea
The similar expression for the
correlator $\langle\Ups_1(z_1;0)\Ups_2(z_2;\thh_2)\rangle$
is obtained from the second one by interchanging the indices.

\subsection{Three-point conformal blocks}

Based on the ansatz
\bea
 &&\langle\Ups_1(z_1;\thh_1)\Ups_2(z_2;\thh_2)
  \Ups_3(z_3;\thh_3)\rangle\nn
 &=&
  \left\{\Ah(\thh_1,\thh_2,\thh_3)+
   \Bh_{1}(\thh_1,\thh_2,\thh_3)\ln z_{12}
   +\Bh_{2}(\thh_1,\thh_2,\thh_3)\ln z_{23}
   +\Bh_{3}(\thh_1,\thh_2,\thh_3)\ln z_{13}\right.\nn
 &&+\ \Dha_{11}(\thh_1,\thh_2,\thh_3)\ln^2 z_{12}
  +\Dha_{12}(\thh_1,\thh_2,\thh_3)\ln z_{12}\ln z_{23}
  +\Dha_{13}(\thh_1,\thh_2,\thh_3)\ln z_{12}\ln z_{13}\nn
  &&+\left. \Dha_{22}(\thh_1,\thh_2,\thh_3)\ln^2 z_{23}
  +\Dha_{23}(\thh_1,\thh_2,\thh_3)\ln z_{23}\ln z_{13}
  +\Dha_{33}(\thh_1,\thh_2,\thh_3)\ln^2 z_{13}\right\}
 z_{12}^{-h_{1}}z_{23}^{-h_{2}}z_{13}^{-h_{3}}\nn
\label{3ans}
\eea
where
\be
 \Ah(\thh_1,\thh_2,\thh_3)\ =\ 
  \Ah^0+\Ah^1\thh_1+\Ah^2\thh_2+\Ah^3\thh_3
  +\Ah^{12}\thh_1\thh_2+\Ah^{23}\thh_2\thh_3+\Ah^{13}\thh_1\thh_3
  +\Ah^{123}\thh_1\thh_2\thh_3
\label{Athh}
\ee
and similarly for $\Bh_{i}(\thh_1,\thh_2,\thh_3)$ and 
$\Dha_{ij}(\thh_1,\thh_2,\thh_3)$,
the general (generating-function) three-point conformal blocks read
\bea
 &&\langle\Ups_1(z_1;0)\Ups_2(z_2;0)\Ups_3(z_3;0)
  \rangle\ =\ \Ah^0V_3\nn
 &&\langle\Ups_1(z_1;\thh_1)\Ups_2(z_2;0)\Ups_3(z_3;0)
  \rangle\ =\ \left\{\Ah^0+\Ah^1\thh_1
   -\Ah^0\thh_1\ln\frac{z_{12}z_{13}}{z_{23}}\right\}V_3\nn  &&\langle\Ups_1(z_1;\thh_1)\Ups_2(z_2;\thh_2)\Ups_3(z_3;0)
  \rangle\nn 
 &&\ \ \ \ =\ \left\{\Ah^0+\Ah^1\thh_1-\Ah^0\thh_1\ln\frac{z_{12}z_{13}}{z_{23}}
  +\Ah^2\thh_2-\Ah^0\thh_2\ln\frac{z_{12}z_{23}}{z_{13}}\right.\nn
 &&\ \ \ \ \ \ \ \ \ +\left. \Ah^{12}\thh_1\thh_2
    -\Ah^1\thh_1\thh_2\ln\frac{z_{12}z_{23}}{z_{13}}
   -\Ah^2\thh_1\thh_2\ln\frac{z_{12}z_{13}}{z_{23}}
   +\Ah^0\thh_1\thh_2\ln\frac{z_{12}z_{23}}{z_{13}}
   \ln\frac{z_{12}z_{13}}{z_{23}}\right\}V_3\nn
&&\langle\Ups_1(z_1;\thh_1)\Ups_2(z_2;\thh_2)
    \Ups_3(z_3;\thh_3)\rangle\nn 
&&\ \ \ \ =\ \left\{\Ah^1\thh_1+\Ah^2\thh_2+\Ah^3\thh_3+\Ah^{12}\thh_1\thh_2
   -\Ah^1\thh_1\thh_2\ln\frac{z_{12}z_{23}}{z_{13}}
   -\Ah^2\thh_1\thh_2\ln\frac{z_{12}z_{13}}{z_{23}}\right.\nn
 &&\ \ \ \ \ \ \ \ \ +\ \Ah^{23}\thh_2\thh_3
  -\Ah^2\thh_2\thh_3\ln\frac{z_{23}z_{13}}{z_{12}}
  -\Ah^3\thh_2\thh_3\ln\frac{z_{13}}{z_{12}z_{23}}
  +\Ah^{13}\thh_1\thh_3
  -\Ah^1\thh_1\thh_3\ln\frac{z_{23}z_{13}}{z_{12}}\nn
&&\ \ \ \ \ \ \ \ \ -\ \Ah^3\thh_1\thh_3\ln\frac{z_{12}z_{13}}{z_{23}}
  +\Ah^{123}\thh_1\thh_2\thh_3
  -\Ah^{12}\thh_1\thh_2\thh_3\ln\frac{z_{23}z_{13}}{z_{12}}\nn
 &&\ \ \ \ \ \ \ \ \ -\ \Ah^{23}\thh_1\thh_2\thh_3\ln\frac{z_{12}z_{13}}{z_{23}}
  -\Ah^{13}\thh_1\thh_2\thh_3\ln\frac{z_{12}z_{23}}{z_{13}}
  +\Ah^1\thh_1\thh_2\thh_3
  \ln\frac{z_{23}z_{12}}{z_{13}}\ln\frac{z_{23}z_{13}}{z_{12}}\nn
 &&\ \ \ \ \ \ \ \ \ +\ \Ah^2\thh_1\thh_2\thh_3
  \ln\frac{z_{12}z_{13}}{z_{23}}\ln\frac{z_{23}z_{13}}{z_{12}}
   +\left. \Ah^3\thh_1\thh_2\thh_3
  \ln\frac{z_{12}z_{23}}{z_{13}}\ln\frac{z_{12}z_{13}}{z_{23}}
  \right\}V_3
\label{3}
\eea
Here we have introduced the abbreviation
\be
 V_3\ =\ \frac{1}{z_{12}^{\Dh_1+\Dh_2-\Dh_3}z_{23}^{-\Dh_1+\Dh_2+\Dh_3}
   z_{13}^{\Dh_1-\Dh_2+\Dh_3}}
\label{V3}
\ee
The remaining correlators are obtained by appropriate permutations in
the indices.

As in the case of two-point conformal blocks, the three-point conformal blocks
may be represented in terms of generalized conformal weights,
$\Dh_i+\thh_i$:
\bea
&&\langle\Ups_1(z_1;0)\Ups_2(z_2;0)\Ups_3(z_3;0)
  \rangle\ =\ \frac{\Ah^0}{z_{12}^{\Dh_1+\Dh_2-\Dh_3}
   z_{23}^{-\Dh_1+\Dh_2+\Dh_3}z_{13}^{\Dh_1-\Dh_2+\Dh_3}}\nn
 &&\langle\Ups_1(z_1;\thh_1)\Ups_2(z_2;0)\Ups_3(z_3;0)
 \rangle\ =\ \frac{\Ah^0+\Ah^1\thh_1}{
   z_{12}^{(\Dh_1+\thh_1)+\Dh_2-\Dh_3}z_{23}^{-(\Dh_1+\thh_1)+\Dh_2+\Dh_3}
   z_{13}^{(\Dh_1+\thh_1)-\Dh_2+\Dh_3}}\nn
 &&\langle\Ups_1(z_1;\thh_1)\Ups_2(z_2;\thh_2)\Ups_3(z_3;0)
  \rangle\nn 
  &&\ \ \ \ =\ \frac{\Ah^0+\Ah^1\thh_1+\Ah^2\thh_2+\Ah^{12}\thh_1\thh_2}{
   z_{12}^{(\Dh_1+\thh_1)+(\Dh_2+\thh_2)-\Dh_3}
   z_{23}^{-(\Dh_1+\thh_1)+(\Dh_2+\thh_2)+\Dh_3}
   z_{13}^{(\Dh_1+\thh_1)-(\Dh_2+\thh_2)+\Dh_3}} \nn
 &&\langle\Ups_1(z_1;\thh_1)\Ups_2(z_2;\thh_2)
    \Ups_3(z_3;\thh_3)\rangle\nn 
 &&\ \ \ \ =\ \frac{\Ah^1\thh_1+\Ah^2\thh_2+\Ah^3\thh_3
   +\Ah^{12}\thh_1\thh_2
   +\Ah^{23}\thh_2\thh_3+\Ah^{13}\thh_1\thh_3
   +\Ah^{123}\thh_1\thh_2\thh_3}{
   z_{12}^{(\Dh_1+\thh_1)+(\Dh_2+\thh_2)-(\Dh_3+\thh_3)}
   z_{23}^{-(\Dh_1+\thh_1)+(\Dh_2+\thh_2)+(\Dh_3+\thh_3)}
   z_{13}^{(\Dh_1+\thh_1)-(\Dh_2+\thh_2)+(\Dh_3+\thh_3)}} 
\label{3th}
\eea
The remaining four combinations are obtained by appropriate
permutations in the indices.

\subsection{Hierarchical structures for conformal blocks}
\label{derive}

Based on ideas discussed in \cite{FloOPE,RAK}, it was found in \cite{logWard}
that the correlators involving logarithmic fields may be represented
as follows:
\bea
  \langle\Psi_1(z_1)\Ups_2(z_2;0)\rangle&=&\Ah^1V_2\nn
  \langle\Psi_1(z_1)\Phi_2(z_2)\rangle&=&\Ah^1V_2\nn
  \langle\Psi_1(z_1)\Psi_2(z_2)\rangle&=&\left(\Ah^{12}+\Ah^2\pa_{\Dh_1}
    +\Ah^1\pa_{\Dh_2}\right)V_2\nn
   \langle\Psi_1(z_1)\Ups_2(z_2;0)
   \Ups_3(z_3;0)\rangle&=&\left(\Ah^1+\Ah^0\pa_{\Dh_1}\right)V_3\nn
 \langle\Psi_1(z_1)\Phi_2(z_2)
   \Ups_3(z_3;0)\rangle&=&\left(\Ah^1+\Ah^0\pa_{\Dh_1}\right)V_3\nn
 \langle\Psi_1(z_1)\Psi_2(z_2)
   \Ups_3(z_3;0)\rangle&=&\left(\Ah^{12}+\Ah^1\pa_{\Dh_2}
    +\Ah^2\pa_{\Dh_1}+A^0\pa_{\Dh_1}\pa_{\Dh_2}\right)V_3\nn
   \langle\Psi_1(z_1)\Phi_2(z_2)
   \Phi_3(z_3)\rangle&=&\Ah^1V_3\nn
 \langle\Psi_1(z_1)\Psi_2(z_2)
   \Phi_3(z_3)\rangle&=&\left(\Ah^{12}+\Ah^2\pa_{\Dh_1}
    +\Ah^1\pa_{\Dh_2}\right)V_3\nn
 \langle\Psi_1(z_1)\Psi_2(z_2)
   \Psi_3(z_3)\rangle&=&
  \left(\Ah^{123}+\Ah^{23}\pa_{\Dh_1}+\Ah^{13}\pa_{\Dh_2}
   +\Ah^{12}\pa_{\Dh_3}\right.\nn
  &&+\left. \Ah^3\pa_{\Dh_1}\pa_{\Dh_2}+\Ah^1\pa_{\Dh_2}\pa_{\Dh_3}
   +\Ah^2\pa_{\Dh_1}\pa_{\Dh_3}\right)V_3
\label{3D}
\eea 
in addition to expressions obtained by appropriately permuting
the indices.
One may therefore represent the correlators hierarchically as
\bea
 \langle\Psi_1(z_1)\Ups_2(z_2;0)\rangle&=&\Ah^1V_2
   +\pa_{\Dh_1}\langle\Phi_1(z_1)\Ups_2(z_2;0)\rangle\nn
  \langle\Psi_1(z_1)\Phi_2(z_2)\rangle&=&\Ah^1V_2
    +\pa_{\Dh_1}\langle\Phi_1(z_1)\Phi_2(z_2)\rangle\nn
  \langle\Psi_1(z_1)\Psi_2(z_2)\rangle&=&\Ah^{12}V_2
   +\pa_{\Dh_1}\langle\Phi_1(z_1)\Psi_2(z_2)\rangle
    +\pa_{\Dh_2}\langle\Psi_1(z_1)\Phi_2(z_2)\rangle
  -\pa_{\Dh_1}\pa_{\Dh_2}\langle\Phi_1(z_1)\Phi_2(z_2)\rangle\nn
\label{2h}
\eea
in the case of two-point conformal blocks, and
\bea
   \langle\Psi_1(z_1)\Ups_2(z_2;0)
   \Ups_3(z_3;0)\rangle&=&\Ah^1V_3
   +\pa_{\Dh_1}\langle\Phi_1(z_1)\Ups_2(z_2;0)
   \Ups_3(z_3;0)\rangle\nn
 \langle\Psi_1(z_1)\Phi_2(z_2)
   \Ups_3(z_3;0)\rangle&=&\Ah^1V_3
   +\pa_{\Dh_1}\langle\Phi_1(z_1)\Phi_2(z_2)
   \Ups_3(z_3;0)\rangle\nn
 \langle\Psi_1(z_1)\Psi_2(z_2)
   \Ups_3(z_3;0)\rangle&=&\Ah^{12}V_3
    +\pa_{\Dh_1}\langle\Phi_1(z_1)\Psi_2(z_2)
      \Ups_3(z_3;0)\rangle
  +\pa_{\Dh_2}\langle\Psi_1(z_1)\Phi_2(z_2)
      \Ups_3(z_3;0)\rangle\nn
  &-&\pa_{\Dh_1}\pa_{\Dh_2}\langle\Phi_1(z_1)\Phi_2(z_2)
    \Ups_3(z_3;0)\rangle\nn
   \langle\Psi_1(z_1)\Phi_2(z_2)
   \Phi_3(z_3)\rangle&=&\Ah^1V_3
   +\pa_{\Dh_1}\langle\Phi_1(z_1)\Phi_2(z_2)
   \Phi_3(z_3)\rangle\nn
 \langle\Psi_1(z_1)\Psi_2(z_2)
   \Phi_3(z_3)\rangle&=&\Ah^{12}V_3
    +\pa_{\Dh_1}\langle\Phi_1(z_1)\Psi_2(z_2)
      \Phi_3(z_3)\rangle
  +\pa_{\Dh_2}\langle\Psi_1(z_1)\Phi_2(z_2)
      \Phi_3(z_3)\rangle\nn
  &-&\pa_{\Dh_1}\pa_{\Dh_2}\langle\Phi_1(z_1)\Phi_2(z_2)
    \Phi_3(z_3)\rangle\nn
 \langle\Psi_1(z_1)\Psi_2(z_2)
   \Psi_3(z_3)\rangle&=&\Ah^{123}V_3
   +\pa_{\Dh_1}\langle\Phi_1(z_1)\Psi_2(z_2)
   \Psi_3(z_3)\rangle
  +\pa_{\Dh_2}\langle\Psi_1(z_1)\Phi_2(z_2)
   \Psi_3(z_3)\rangle\nn
 &+&\pa_{\Dh_3}\langle\Psi_1(z_1)\Psi_2(z_2)
   \Phi_3(z_3)\rangle
  -\pa_{\Dh_1}\pa_{\Dh_2}\langle\Phi_1(z_1)\Phi_2(z_2)
   \Psi_3(z_3)\rangle\nn
  &-&\pa_{\Dh_2}\pa_{\Dh_3}\langle\Psi_1(z_1)\Phi_2(z_2)
   \Phi_3(z_3)\rangle
  -\pa_{\Dh_1}\pa_{\Dh_3}\langle\Phi_1(z_1)\Psi_2(z_2)
   \Phi_3(z_3)\rangle\nn
 &+&\pa_{\Dh_1}\pa_{\Dh_2}\pa_{\Dh_3}\langle\Phi_1(z_1)\Phi_2(z_2)
   \Phi_3(z_3)\rangle
\label{3h}
\eea
in the case of three-point conformal blocks. As above, the remaining correlators 
may be obtained by appropriately permuting the indices.

\section{On $SL(2,\mathbb{R})$ WZW models}
\label{WZW}

The affine $sl(2)_k$ Lie algebra, 
including the commutators with the Virasoro modes, reads
\bea
 \left[J_{+,n},J_{-,m}\right]&=&2J_{0,n+m}+kn\delta_{n+m,0} \nn
 \left[J_{0,n},J_{\pm,m}\right]&=&\pm J_{\pm,n+m} \nn
 \left[J_{0,n},J_{0,m}\right]&=&\frac{k}{2}n\delta_{n+m,0} \nn
 \left[L_n,J_{a,m}\right]&=&-mJ_{a,n+m}
\label{aff}
\eea
Another conventional notation is obtained by replacing 
$\{J_{+,n},2J_{0,n},J_{-,n}\}$
by $\{E_n,H_n,F_n\}$.
The level of the algebra is indicated by $k$ and 
is related to the central charge as $c=3k/(k+2)$.
The non-vanishing entries of the Cartan-Killing form of $sl(2)$ 
are given by  
\be
 \kappa_{00}=\hf,\ \ \ \ \ \ \ \ \kappa_{+-}=\kappa_{-+}=1
\label{CK}
\ee
and appear as coefficients to the central terms in (\ref{aff}).
Its inverse is given by
\be
 \kappa^{00}=2,\ \ \ \ \ \ \ \ \kappa^{+-}=\kappa^{-+}=1
\label{CKinv}
\ee
and comes into play when discussing the affine Sugawara construction below.
We will be concerned mainly with the `horizontal'
part of the affine Lie algebra, the $sl(2)$ Lie algebra generated
by the zero modes $\{J_a:=J_{a,0}\}$.

We will assume that the 
Virasoro primary fields of a given conformal weight $\D$ may be
organized in multiplets corresponding to spin-$j$ 
representations of the $sl(2)$ algebra, where 
\be
 \D=\frac{j(j+1)}{k+2}
\label{Dj}
\ee
In the following, $j$ is taken to be real even though the general
formalism is amenable to treat $j$ complex as well.
If $2j$ is a non-negative integer, we may
label the $2j+1$ members of the associated multiplet as in
\be
 \phi_{-j}(z),\ \phi_{-j+1}(z),\ \dots,\ \phi_{j-1}(z),\ \phi_{j}(z)
\label{phij}
\ee
where the dependence on $j$, often indicated by
$\phi_{j,m}$, is suppressed. A finite-dimensional representation
like (\ref{phij}) is often referred to as an integrable representation.
The field $\phi_{m}$ has $J_{0}$ eigenvalue $m$, while we will
use the following convenient
choice of relative normalizations of the fields:
\bea
 \left[J_+,\phi_m(z)\right]&=&(j+m+1)\phi_{m+1}(z)\nn
 \left[J_0,\phi_m(z)\right]&=&m\phi_{m}(z)\nn
 \left[J_-,\phi_m(z)\right]&=&(j-m+1)\phi_{m-1}(z)
\label{Jphi}
\eea
If $2j$ is not a non-negative integer, the associated
primary fields may in general 
be organized in an infinite-dimensional multiplet corresponding to an
$sl(2)$ representation.

\subsection{Generating-function primary fields}

A generating function for the $2j+1$ Virasoro primary fields in an
integrable representation may be written \cite{FZ}
\be
 \phi(z,x)\ =\ \sum_{m=-j}^j\phi_m(z)x^{j-m}
\label{phisum}
\ee
To keep the notation simple here and in the following, we do not 
indicate explicitly whether the sum is over integers or half-integers as 
this should be obvious from the integer or half-integer nature of the spin itself.
For general spin and associated infinite-dimensional multiplet,
the generating function for a so-called highest-weight representation, for
example, reads
\be
 \phi(z,x)\ =\ \sum_{m\in(j+\mathbb{Z}_{\leq})}\phi_m(z)x^{j-m}
\label{gfhw}
\ee
The adjoint action of the affine generators on the generating-function
primary field reads
\be
 [J_{a,n},\phi(z,x)]\ =\ -z^nD_a(x)\phi(z,x)
\label{Jphix}
\ee
where the differential operators $D_a(x)$ are defined by
\bea
 D_+(x)&=&x^2\pa_x-2jx\nn
 D_0(x)&=&x\pa_x-j\nn
 D_-(x)&=&-\pa_x
\label{D}
\eea
They generate the Lie algebra $sl(2)$, and one recovers (\ref{Jphi})
from (\ref{Jphix}).

A correlator like the ${\cal N}$-point chiral block
\be
 \langle\phi_1(z_1,x_1)\dots\phi_{\cal N}(z_{\cal N},x_{\cal N})\rangle
\label{Nzx}
\ee
is seen to correspond to a generating function for the individual
correlators based on fields, $\phi_{i,m_i}(z_i)$, appearing
in expansions like (\ref{phisum}) (or (\ref{gfhw}), for example).
That is, the ${\cal N}$-point chiral block
\be
  \langle\phi_{1,m_1}(z_1)\dots
   \phi_{{\cal N},m_{\cal N}}(z_{\cal N})\rangle
\label{Nz}
\ee
appears as the coefficient to $\prod_{i=1}^{\cal N}x_i^{j_i-m_i}$
in an expansion of (\ref{Nzx}). The general expansion thus reads
\be
 \langle\phi_1(z_1,x_1)\dots\phi_{\cal N}(z_{\cal N},x_{\cal N})\rangle
 \ =\ \sum_{m_1,\dots,m_{\cal N}}
   \langle\phi_{1,m_1}(z_1)\dots
   \phi_{{\cal N},m_{\cal N}}(z_{\cal N})\rangle x_1^{j_1-m_1}\dots 
   x_{\cal N}^{j_{\cal N}-m_{\cal N}}
\label{Nzexp}
\ee
where the ranges of the summation variables depend on the 
individual spin-$j_i$ representations.

\subsection{The KZ equations}

In a WZW model, the Virasoro generators are
realized as bilinear expressions in the affine generators.
This is referred to as the affine Sugawara construction
which is here written in terms of modes
\be
  L_N\ =\ 
   \frac{1}{2(k+2)}\kappa^{ab}\left(\sum_{n\leq-1}J_{a,n}J_{b,N-n}
  +\sum_{n\geq0}J_{a,N-n}J_{b,n}\right)
\label{Sug}
\ee
Here and in the following, we will use the convention of
summing over appropriately repeated group indices, $a=\pm,0$.
Acting on a highest-weight state, the affine Sugawara
construction gives rise to singular vectors of the combined
algebra. The decoupling of these is trivial for $N>0$. For $N=0$,
it reproduces the relation (\ref{Dj}) as the eigenvalue of $L_0$
is equated with the eigenvalue of the normalized quadratic Casimir:
\be
 \D\ =\ \frac{\kappa^{ab}D_a(x)D_b(x)}{2(k+2)}\ =\ \frac{j(j+1)}{k+2}
\label{Cas}
\ee 

The condition corresponding to $N=-1$ leads to the
celebrated KZ equations \cite{KZ} which are written here
for an ${\cal N}$-point chiral block of generating-function primary fields
\be
 0\ =\ KZ_i\langle\phi_1(z_1,x_1)\dots
  \phi_{\cal N}(z_{\cal N},x_{\cal N})\rangle,\ \ \ \ \ \ \ i=1,\dots,{\cal N}
\label{KZ}
\ee
where
\be
 KZ_i\ =\ \left((k+2)\pa_{z_i}-\sum_{j\neq i}
  \frac{\kappa^{ab}D_a(x_i)D_b(x_j)}{z_i-z_j}
 \right)
\label{KZi}
\ee
These ${\cal N}$ differential equations associated
to a given ${\cal N}$-point chiral block are not all independent. This is easily
illustrated by considering the sum 
\be
 \sum_{i=1}^{\cal N}KZ_i\ =\ (k+2)\sum_{i=1}^{\cal N}\pa_{z_i}
\label{sumKZ}
\ee
which merely induces translational invariance already imposed
by the first conformal Ward identity (\ref{confward}).
As we will discuss below, a simple modification of the KZ equations
(\ref{KZ}), (\ref{KZi})
apply to correlators involving certain logarithmic fields to be introduced
in the following.

\section{Affine Jordan cells}
\label{affJc}

We wish to consider the situation where
every Virasoro primary field in a given $sl(2)$ representation may 
have a logarithmic partner. The resulting multiplet of fields
is comprised of primary fields as well as so-called logarithmic fields and
will be referred to as an affine Jordan cell.
A priori, the hosting model may consist of a family of affine Jordan cells
in coexistence with an independent family of multiplets of primary fields
without logarithmic partners. We will refer loosely to such a model as a logarithmic
WZW model.
Primary fields not appearing in an affine
Jordan cell will be called non-cellular primary fields.
It is found that the affine Jordan cells relevant to our studies
contain primary fields not having logarithmic partners.
These primary fields are naturally included in the generating
functions for the logarithmic fields rather than in the generating
functions for the primary fields comprising the original
spin-$j$ representation we are extending. 
To reach this appreciation of the affine Jordan cells, 
we initially consider an extension
of the differential-operator realization (\ref{D})
and its role in a generalization of (\ref{Jphix}).

\subsection{Generating-function unified cells}

The differential-operator realization  
\bea
 D_+(x;\theta)&=&x^2\pa_x-2(j+\theta)x\nn
 D_0(x;\theta)&=&x\pa_x-(j+\theta)\nn
 D_-(x;\theta)&=&-\pa_x
\label{Dth}
\eea
of the Lie algebra $sl(2)$
is designed to act on a representation of (generalized) spin $j+\theta$,
where $\theta$ is a nilpotent, yet even, parameter satisfying $\theta^2=0$.
Extending the idea of organizing fields in generating functions
as in (\ref{phisum}) satisfying (\ref{Jphix}), we introduce the formal
generating-function unified cell $\Ups(z,x;\theta)$ satisfying
\be
 [J_a,\Ups(z,x;\theta)]\ =\ -D_a(x;\theta)\Ups(z,x;\theta)
\label{JDUps}
\ee
We note that this also applies to generating-function primary
fields as it reduces to (\ref{Jphix}) (for $n=0$) when we set $\theta=0$.
Here and in the following, focus is on the $sl(2)$ Lie algebra part of
the affine generators.
An expansion of the generating-function unified cell 
with respect to $\theta$ may be written
\be
 \Ups(z,x;\theta)\ =\ \Phi(z,x)+\theta\Psi(z,x)\nn
\label{Upsxexp}
\ee
resembling the definition of the unified cell (\ref{Ups}) in logarithmic CFT.
In terms of the new generating functions, $\Phi(z,x)$ and $\Psi(z,x)$, 
the commutators (\ref{JDUps}) read
\bea
 \left[J_+,\Phi(z,x)\right]&=&-D_+(x)\Phi(z,x)\nn
 \left[J_+,\Psi(z,x)\right]&=&-D_+(x)\Psi(z,x)+2x\Phi(z,x)\nn
 \left[J_0,\Phi(z,x)\right]&=&-D_0(x)\Phi(z,x)\nn
 \left[J_0,\Psi(z,x)\right]&=&-D_0(x)\Psi(z,x)+\Phi(z,x)\nn
 \left[J_-,\Phi(z,x)\right]&=&-D_-(x)\Phi(z,x)\nn
 \left[J_-,\Psi(z,x)\right]&=&-D_-(x)\Psi(z,x)
\label{JDPP}
\eea
where the differential operators, $D_a(x)$, are given in (\ref{D}).

These commutators severely restrict the set of $sl(2)$
representations for which the two fields $\Phi(z,x)$ and $\Psi(z,x)$ can be
considered generating functions.
It is beyond the scope of the present work, though, to classify
these representations, even in the simple case where $\Phi(z,x)$
is the generating function for a finite-dimensional 
representation as in (\ref{phisum}). We hope to address 
this classification elsewhere.
Here we merely wish to demonstrate the existence
of representations corresponding to the generating
functions (\ref{Upsxexp}), (\ref{JDPP}) and to illustrate their complexity.
We will do so by considering a particular 
logarithmic extension of a finite-dimensional spin-$j$ representation.
More general examples are considered in Section \ref{genexp}.

We thus introduce the following expansions of the generating
functions $\Phi(z,x)$ and $\Psi(z,x)$:
\be
 \Phi(z,x)\ =\ \sum_{m=-j}^j\Phi_m(z)x^{j-m},\ \ \ \ \ \ \ 
 \Psi(z,x)\ =\ \sum_{m=-\infty}^j\Psi_m(z)x^{j-m}
\label{Phisum}
\ee
The remark following (\ref{phisum}) about $m$ taking on integer
or half-integer values also applies when one (or even both) of the
summation bounds is (either plus or minus) infinity.
As already mentioned, we are concerned with Jordan cells
whose principal parts correspond to finite-dimensional spin-$j$ representations,
here governed by the generating function $\Phi(z,x)$.
It is noted that the logarithmic part, on the other hand, 
consists of infinitely many fields. 

With the understanding that $\Phi_m(z)$ only exists for
$m=-j,\dots,j$ while $\Psi_m(z)$ only exists for
$m=-\infty,\dots,j$, the adjoint action of the $sl(2)$ Lie algebra
on the modes of the two generating functions (\ref{Phisum})
may be written compactly as
\bea
 \left[J_+,\Phi_m(z)\right]&=&(j+m+1)\Phi_{m+1}(z)\nn
 \left[J_+,\Psi_m(z)\right]&=&(j+m+1)\Psi_{m+1}(z)+2\Phi_{m+1}(z)\nn
 \left[J_0,\Phi_m(z)\right]&=&m\Phi_{m}(z)\nn
 \left[J_0,\Psi_m(z)\right]&=&m\Psi_{m}(z)+\Phi_{m}(z)\nn
 \left[J_-,\Phi_m(z)\right]&=&(j-m+1)\Phi_{m-1}(z)\nn
 \left[J_-,\Psi_m(z)\right]&=&(j-m+1)\Psi_{m-1}(z)
\label{JPP}
\eea
This is equivalent to simply setting a non-existing field
equal to zero whenever it formally appears in (\ref{JPP}).
The following diagram may help visualizing the representation:
\be
 \mbox{
 \begin{picture}(100,160)(160,-40)
    \unitlength=1.4cm
  \thinlines
  \put(2,2){$\Psi_{-j-1}$}
  \put(4,0){$\Phi_{-j}$}
   \put(4,2){$\Psi_{-j}$}
  \put(6,0){$\Phi_{-j+1}$}
   \put(6,2){$\Psi_{-j+1}$}
  \put(10,0){$\Phi_{j}$}
   \put(10,2){$\Psi_{j}$}
  \put(1,2){$\longleftrightarrow$}
  \put(3,2){$\longleftarrow$}
 \put(5,0){$\longleftrightarrow$}
  \put(5,2){$\longleftrightarrow$}
  \put(7,0){$\longleftrightarrow$}
  \put(7,2){$\longleftrightarrow$}
  \put(9,0){$\longleftrightarrow$}
  \put(9,2){$\longleftrightarrow$}
  \put(3,1){$\searrow$}
  \put(5,1){$\searrow$}
  \put(7,1){$\searrow$}
  \put(9,1){$\searrow$}
  \put(4,1){$\downarrow$}
  \put(6,1){$\downarrow$}
  \put(10,1){$\downarrow$}
 \put(0,2){$\dots$}
 \put(8,0){$\dots$}
  \put(8,1){$\dots$}
 \put(8,2){$\dots$}
 \put(0.9,2.2){$J_-,J_+$}
 \put(3.05,2.2){$J_-$}
 \put(4.9,2.2){$J_-,J_+$}
 \put(6.9,2.2){$J_-,J_+$}
 \put(8.9,2.2){$J_-,J_+$}
 \put(4.9,-0.25){$J_-,J_+$}
 \put(6.9,-0.25){$J_-,J_+$}
 \put(8.9,-0.25){$J_-,J_+$}
  \put(3.15,1.1){$J_+$}
  \put(5.15,1.1){$J_+$}
  \put(7.15,1.1){$J_+$}
  \put(9.15,1.1){$J_+$}
  \put(4.15,1){$J_0$}
  \put(6.15,1){$J_0$}
  \put(10.15,1){$J_0$}
\end{picture}
}
\label{hw}
\ee
Here the arrows indicate the adjoint actions of $J_a$ (except the 
primary parts of the adjoint actions of $J_0$ which are not indicated explicitly).
It is observed that only $2j+1$ of the fields $\Psi_m(z)$ are logarithmic
fields, where a logarithmic field is characterized by the property that
at least one of the affine generators acts non-diagonally
on them. Here, in particular, they do not have well-defined 
$J_0$ eigenvalues.

The naive expansion where $\Psi(z,x)$ is a sum of $2j+1$
fields similar to the expansion of $\Phi(z,x)$ turns out to
be inconsistent. The same problem occurs when trying to
write $\Psi(z,x)$ as an infinite sum from $-j$ to $\infty$,
as it actually occurs for all power-series expansions of $\Psi(z,x)$
having lowest magnetic moment, $m$, equal to $-j$.
This asymmetry in extensions beyond $j$ and $-j$, respectively,
stems from the fact that $J_+$ may act non-diagonally 
(in the algebraic sense (\ref{JPP}), i.e., diagonally in the diagram
(\ref{hw})) while
$J_-$ only acts diagonally (i.e., horizontally in the diagram (\ref{hw})). 
One can extend in both directions simultaneously
\be
  \Phi(z,x)\ =\ \sum_{m=-j}^j\Phi_m(z)x^{j-m},\ \ \ \ \ \ \ 
 \Psi(z,x)\ =\ \sum_{m=-\infty}^\infty\Psi_m(z)x^{j-m}
\label{Phisum2}
\ee
in which case one obtains the following reducible extension
of (\ref{hw}): 
\be
 \mbox{
 \begin{picture}(100,160)(120,-40)
    \unitlength=1.4cm
  \thinlines
   \put(0,2){$\Psi_{-j-1}$}
  \put(2,0){$\Phi_{-j}$}
   \put(2,2){$\Psi_{-j}$}
  \put(6,0){$\Phi_{j}$}
   \put(7.5,2){$\Psi_{j+1}$}
 \put(6,2){$\Psi_{j}$}
 \put(-0.7,2){$\longleftrightarrow$}
  \put(1,2){$\longleftarrow$}
   \put(3,0){$\longleftrightarrow$}
  \put(3,2){$\longleftrightarrow$}
 \put(5,0){$\longleftrightarrow$}
  \put(5,2){$\longleftrightarrow$}
  \put(6.7,2){$\longrightarrow$}
  \put(8.3,2){$\longleftrightarrow$}
  \put(1,1){$\searrow$}
  \put(3,1){$\searrow$}
  \put(5,1){$\searrow$}
 \put(2,1){$\downarrow$}
  \put(6,1){$\downarrow$}
 \put(4,0){$\dots$}
 \put(4,1){$\dots$}
 \put(4,2){$\dots$}
 \put(-1.2,2){$\dots$}
 \put(9.1,2){$\dots$}
 \put(1.05,2.2){$J_-$}
 \put(2.9,2.2){$J_-,J_+$}
 \put(4.9,2.2){$J_-,J_+$}
 \put(6.75,2.2){$J_+$}
 \put(8.2,2.2){$J_-,J_+$}
 \put(2.9,-0.25){$J_-,J_+$}
 \put(4.9,-0.25){$J_-,J_+$}
  \put(1.15,1.1){$J_+$}
  \put(3.15,1.1){$J_+$}
  \put(5.15,1.1){$J_+$}
 \put(2.15,1){$J_0$}
  \put(6.15,1){$J_0$}
\end{picture}
}
\label{cont}
\ee
The representation (\ref{hw}) is obtained from (\ref{cont})
by factoring out the submodule generated from $\Psi_{j+1}$.
The form of the structure constants in (\ref{JPP}) allow us to
indicate both representations by the same commutator algebra (\ref{JPP})
and in both cases write the expansion of $\Psi(z,x)$ as a sum
over all $m$. Infinitely many terms will be redundant, though,
when writing the expansion corresponding to (\ref{hw}) in this way.
Other representations can be envisaged (cf. Section \ref{genexp}), 
and as already indicated, we hope to return elsewhere
with a discussion of the classification of affine Jordan cells defined
as logarithmic (i.e., non-diagonal or indecomposable) 
extensions of integrable or non-integrable 
(affine) $sl(2)$ representations.

In the two examples discussed above, (\ref{Phisum}) and (\ref{Phisum2}), 
the generating function for the unified 
cell (\ref{Upsxexp}) may be expanded as
\be
 \Ups(z,x;\theta)\ =\ \sum_m\Ups_m(z;\theta)x^{j-m}\ =\ 
  \sum_m\Phi_m(z)x^{j-m}
  +\theta\sum_m\Psi_m(z)x^{j-m}
\label{UpsPPx}
\ee
where the ranges for the summation variables may be different
in the last two sums, cf. (\ref{Phisum2}), for example.
Analogous to the discussion of unified cells following (\ref{Lup}),
we reserve the notation $\Upsilon(z,x;0)$ for generating-function
primary fields not belonging to a unified cell like (\ref{UpsPPx}).
In terms of the modes of the generating-function Jordan cell 
given in (\ref{UpsPPx}), the commutators (\ref{JPP}) read
\bea
  \left[J_+,\Ups_m(z;\theta)\right]&=&(j+m+1+2\theta)\Ups_{m+1}(z;\theta)\nn
 \left[J_0,\Ups_m(z;\theta)\right]&=&(m+\theta)\Ups_{m}(z;\theta)\nn
 \left[J_-,\Ups_m(z;\theta)\right]&=&(j-m+1)\Ups_{m-1}(z;\theta)
\label{JUps}
\eea
As in (\ref{Lup}) where $\Dh+\thh$ may be interpreted as a generalized
conformal weight, we now have a generalized spin and associated
generalized magnetic moments given by $j+\theta$ and $m+\theta$, respectively.
This was already indicated following (\ref{Dth}).

It is recalled that a correlator of generating functions like
(\ref{Nzx}) may be regarded as a generating function for
the individual conformal blocks, cf. (\ref{Nzexp}). 
This principle extends to ${\cal N}$-point chiral blocks involving
generating-function unified cells.
If all ${\cal N}$ generating functions may be expanded as
in (\ref{UpsPPx}), we then have
\bea
 &&\langle\Ups_1(z_1,x_1;\theta_1)\dots
  \Ups_{\cal N}(z_{\cal N},x_{\cal N};\theta_{\cal N})\rangle\nn
 &&\ \ \ \ =\ \sum_{m_1,\dots,m_{\cal N}}
  \langle\Ups_1(z_1;\theta_1)\dots
  \Ups_{\cal N}(z_{\cal N};\theta_{\cal N})\rangle
  x_1^{j_1-m_1}\dots x_{\cal N}^{j_{\cal N}-m_{\cal N}}\nn
 &&\ \ \ \ =\ \sum_{m_1,\dots,m_{\cal N}}\left\{
  \langle\Phi_1(z_1)\dots\Phi_{\cal N}(z_{\cal N})\rangle
  +\theta_1\langle\Psi_1(z_1)\Phi_2(z_2)
    \dots\Phi_{\cal N}(z_{\cal N})\rangle+\dots\right.\nn
 &&\ \ \ \ \ \ \ \ \ \ \ \ \ \ \ \ \ \ \ \ \ \ \ \
  \left.+\ \theta_1\dots\theta_{\cal N}
  \langle\Psi_1(z_1)\dots\Psi_{\cal N}(z_{\cal N})\rangle\right\}
  x_1^{j_1-m_1}\dots x_{\cal N}^{j_{\cal N}-m_{\cal N}}
\label{UpsNexp}
\eea
where $\Ups_i(z_i,x_i;\theta_i)$ denotes a generating-function
primary field if $\theta_i=0$.
As above, the ranges of the summation variables depend on the 
individual $sl(2)$ representations.

\subsection{More on indecomposable $sl(2)$ representations}
\label{genexp}

It is stressed that the differential-operator realization
(\ref{Dth}) and the generating-function (\ref{UpsPPx}) do not exhaust
all possible extensions of the ordinary WZW model 
outlined in Section \ref{WZW}. This is illustrated by the models
discussed in \cite{Gabsu2,FFHST,LMRS,Nic,Nicsu2}, for example, 
and will be addressed further elsewhere.
The construction developed in the present work
has the virtue that it, under hamiltonian reduction, reduces to
the non-affine logarithmic CFT reviewed in Section \ref{LogCFT}.
This will be the topic of Section \ref{hamred} below.
Here we wish to indicate the level of complexity of the $sl(2)$ 
representations associated to more general expansions 
of $\Psi(z,x)$ than power-series expansions such as
(\ref{Phisum2}).

To this end, we consider the expansions
\be
 \Phi(z,x)\ =\ \sum_m\Phi_m(z)x^{j-m},\ \ \ \ \ \ \ 
 \Psi(z,x)\ =\ \sum_{m,n}\Psi_{m,n}(z)x^{j-m}\ln^nx
\label{Psiln}
\ee
where we have left the summation ranges unspecified.
In terms of these modes, the commutators (\ref{JDPP}) read
\bea
 \left[J_+,\Phi_m(z)\right]&=&(j+m+1)\Phi_{m+1}(z)\nn
 \left[J_+,\Psi_{m,n}(z)\right]&=&(j+m+1)\Psi_{m+1,n}(z)
  -(n+1)\Psi_{m+1,n+1}(z)+2\Phi_{m+1}(z)\nn
 \left[J_0,\Phi_m(z)\right]&=&m\Phi_{m}(z)\nn
  \left[J_0,\Psi_{m,n}(z)\right]&=&m\Psi_{m+1,n}(z)
  -(n+1)\Psi_{m,n+1}(z)+\Phi_{m}(z)\nn
  \left[J_-,\Phi_m(z)\right]&=&(j-m+1)\Phi_{m-1}(z)\nn
 \left[J_-,\Psi_{m,n}(z)\right]&=&(j-m+1)\Psi_{m-1,n}(z)
  +(n+1)\Psi_{m-1,n+1}(z)
\label{JPPln}
\eea
where a field whose indices do not match the expansion
(\ref{Psiln}) is set to zero.
To illustrate such an indecomposable $sl(2)$ representation, we let
$n$ run from 0 to 2 and focus on a typical sequence
in the magnetic moments: $m-1$, $m$, $m+1$, where $-j<m<j$.
The announced part of the corresponding diagram then looks like
\be
 \mbox{
 \begin{picture}(100,320)(90,-40)
    \unitlength=1.4cm
  \thinlines
   \put(0,0){$\Phi_{m-1}$}
  \put(0,2){$\Psi_{m-1,0}$}
   \put(0,4){$\Psi_{m-1,1}$}
  \put(0,6){$\Psi_{m-1,2}$}
   \put(3,0){$\Phi_{m}$}
  \put(3,2){$\Psi_{m,0}$}
   \put(3,4){$\Psi_{m,1}$}
  \put(3,6){$\Psi_{m,2}$}
   \put(6,0){$\Phi_{m+1}$}
  \put(6,2){$\Psi_{m+1,0}$}
   \put(6,4){$\Psi_{m+1,1}$}
  \put(6,6){$\Psi_{m+1,2}$}
  \put(1.5,0){$\longleftrightarrow$}
  \put(4.5,0){$\longleftrightarrow$}
 \put(1.5,2){$\longleftrightarrow$}
  \put(4.5,2){$\longleftrightarrow$}
 \put(1.5,4){$\longleftrightarrow$}
  \put(4.5,4){$\longleftrightarrow$}
 \put(1.5,6){$\longleftrightarrow$}
  \put(4.5,6){$\longleftrightarrow$}
  \put(1.5,1){$\searrow$}
  \put(4.5,1){$\searrow$}
  \put(1.5,3){$\nearrow$}
  \put(4.5,3){$\nearrow$}
  \put(1.5,5){$\nearrow$}
  \put(4.5,5){$\nearrow$}
  \put(1.5,3){$\nwarrow$}
  \put(4.5,3){$\nwarrow$}
  \put(1.5,5){$\nwarrow$}
  \put(4.5,5){$\nwarrow$}
 \put(0,1){$\downarrow$}
 \put(3,1){$\downarrow$}
 \put(6,1){$\downarrow$}
 \put(0,3){$\uparrow$}
 \put(3,3){$\uparrow$}
 \put(6,3){$\uparrow$}
 \put(0,5){$\uparrow$}
 \put(3,5){$\uparrow$}
 \put(6,5){$\uparrow$}
  \put(-0.5,2){\line(1,3){0.4}}
  \put(-0.5,2){\line(1,-3){0.4}}
  \put(-0.1,0.8){\line(0,1){0.07}}
  \put(-0.1,0.8){\line(-1,0){0.07}}
  \put(2.5,2){\line(1,3){0.4}}
  \put(2.5,2){\line(1,-3){0.4}}
  \put(2.9,0.8){\line(0,1){0.07}}
  \put(2.9,0.8){\line(-1,0){0.07}}
  \put(5.5,2){\line(1,3){0.4}}
  \put(5.5,2){\line(1,-3){0.4}}
  \put(5.9,0.8){\line(0,1){0.07}}
  \put(5.9,0.8){\line(-1,0){0.07}}
  \put(-0.5,3){\line(1,6){0.4}}
  \put(-0.5,3){\line(1,-6){0.4}}
  \put(-0.1,0.6){\line(0,1){0.07}}
  \put(-0.1,0.6){\line(-1,0){0.07}}
  \put(2.5,3){\line(1,6){0.4}}
  \put(2.5,3){\line(1,-6){0.4}}
  \put(2.9,0.6){\line(0,1){0.07}}
  \put(2.9,0.6){\line(-1,0){0.07}}
  \put(5.5,3){\line(1,6){0.4}}
  \put(5.5,3){\line(1,-6){0.4}}
  \put(5.9,0.6){\line(0,1){0.07}}
  \put(5.9,0.6){\line(-1,0){0.07}}
 \put(0.5,3.5){\line(2,-3){2}}
 \put(0.5,5.5){\line(2,-5){2}}
 \put(2.5,0.5){\line(0,1){0.07}}
 \put(2.5,0.5){\line(-1,0){0.07}}
 \put(3.5,3.5){\line(2,-3){2}}
 \put(3.5,5.5){\line(2,-5){2}}
 \put(5.5,0.5){\line(0,1){0.07}}
 \put(5.5,0.5){\line(-1,0){0.07}}
 \put(-1,0){$\dots$}
 \put(7.3,0){$\dots$}
 \put(-1,2){$\dots$}
 \put(7.3,2){$\dots$}
 \put(-1,4){$\dots$}
 \put(7.3,4){$\dots$}
 \put(-1,6){$\dots$}
 \put(7.3,6){$\dots$}
\end{picture}
}
\label{gen}
\ee
As in (\ref{hw}) and (\ref{cont}), 
the arrows indicate the adjoint actions of the $sl(2)$ generators.
The kinked arrows refer to parts of the $J_0$ actions.

\subsection{Modified KZ equations}

The logarithmic WZW model hosting the affine Jordan cells introduced
above, is based on an extension of the affine Sugawara
construction (\ref{Sug}). As the actions of the Virasoro modes 
depend on the target field being a unified cell or not
(compare (\ref{Lup}) to the first commutator in (\ref{L})), 
the actions of the affine modes appearing in the extended affine 
Sugawara construction must have
a similar dependence. The generalization of (\ref{Cas}) thus reads
\be
 \D+\mu\ =\ \frac{\kappa^{ab}D_a(x;\theta)D_b(x;\theta)}{2(k+2)}\ =\ 
 \frac{(j+\theta)(j+\theta+1)}{k+2}
\label{Casth}
\ee
from which it follows that the nilpotent part of the conformal weight,
$\mu$, is related to the nilpotent part of the spin, $\theta$, as
\be
 \mu\ =\ \frac{2j+1}{k+2}\theta
\label{thhth}
\ee

Likewise, the KZ equations may be extended to cover correlators
of generating-function unified cells simply by replacing $D_a(x_i)$
by $D_a(x_i;\theta_i)$ if the $i$th field is such a unified cell.
With the understanding that a non-cellular generating-function
primary field, $\Ups_i(z_i,x_i;0)$, corresponds to setting $\theta_i=0$ in
$\Ups_i(z_i,x_i;\theta_i)$, the modified KZ equations read
\be
 0\ =\ KZ_i\langle\Ups_1(z_1,x_1;\theta_1)\dots
  \Ups_{\cal N}(z_{\cal N},x_{\cal N};\theta_{\cal N})\rangle,
   \ \ \ \ \ \ \ i=1,\dots,{\cal N}
\label{KZth}
\ee
where
\be
 KZ_i\ =\ \left((k+2)\pa_{z_i}-\sum_{j\neq i}\frac{\kappa^{ab}D_a(x_i;\theta_i)
  D_b(x_j;\theta_j)}{z_i-z_j}\right)
\label{KZith}
\ee
As in the non-logarithmic case, these ${\cal N}$ differential equations
are not all independent as they satisfy (\ref{sumKZ}).
This is true for all combinations of ${\cal N}$ generating-function
fields, i.e., every generating-function field can be a generating-function
primary field or a generating-function Jordan cell.

\section{Correlators in logarithmic $SL(2,\mathbb{R})$ WZW models}
\label{corr}

We now turn to the computation of correlators in the logarithmic
WZW model introduced above. Focus will be on two- and three-point
chiral blocks of generating-functions. 
The correlators are worked out as $SL(2,\mathbb{R})$ 
group-invariant solutions to the conformal
Ward identities and are subsequently demonstrated to
satisfy the generalized KZ equations (\ref{KZth}), (\ref{KZith}).
This means that conformal and $SL(2,\mathbb{R})$ group invariance fix
the form of the two- and three-point chiral blocks as is the case in the
ordinary, non-logarithmic $SL(2,\mathbb{R})$ WZW model.

\subsection{$SL(2,\mathbb{R})$ group invariance and conformal Ward identities}

Bearing the link (\ref{thhth}) in mind, the conformal Ward identities 
(\ref{confward}) now read
\bea
 0&=&\sum_{i=1}^{\cal N}\pa_{z_i}\langle\Ups_1(z_1,x_1;\theta_1)\dots
  \Ups_{\cal N}(z_{\cal N},x_{\cal N};\theta_{\cal N})\rangle\nn
 0&=&\sum_{i=1}^{\cal N}\left(z_i\pa_{z_i}+\D_i+\frac{2j_i+1}{k+2}\theta_i\right)
  \langle\Ups_1(z_1,x_1;\theta_1)\dots
   \Ups_{\cal N}(z_{\cal N},x_{\cal N};\theta_{\cal N})\rangle\nn
 0&=&\left({\cal L}^{\cal N}+2\sum_{i=1}^{\cal N}
  \frac{2j_i+1}{k+2}\theta_iz_i\right)
   \langle\Ups_1(z_1,x_1;\theta_1)\dots
   \Ups_{\cal N}(z_{\cal N},x_{\cal N};\theta_{\cal N})\rangle
\label{confwardx}
\eea
where the differential operator ${\cal L}^{\cal N}$ is defined
in (\ref{calL}). Correlators satisfying these identities are 
said to be projectively invariant. Likewise, invariance under
$SL(2,\mathbb{R})$ group transformations (the ones generated by the horizontal
$sl(2)$ algebra) is sometimes referred to as loop-projective invariance.
The corresponding Ward identities are often called affine Ward
identities, though we will refer to them as $SL(2,\mathbb{R})$ Ward identities. 
For generating-function correlators involving
generating-function Jordan cells (and possibly non-cellular
generating-function primary fields) as the ones appearing
in (\ref{confwardx}), they read
\bea
 0&=&\sum_{i=1}^{\cal N}\pa_{x_i}\langle\Ups_1(z_1,x_1;\theta_1)\dots
  \Ups_{\cal N}(z_{\cal N},x_{\cal N};\theta_{\cal N})\rangle\nn
 0&=&\sum_{i=1}^{\cal N}\left(x_i\pa_{x_i}-j_i-\theta_i\right)
  \langle\Ups_1(z_1,x_1;\theta_1)\dots
   \Ups_{\cal N}(z_{\cal N},x_{\cal N};\theta_{\cal N})\rangle\nn
 0&=&\left({\cal J}^{\cal N}-2\sum_{i=1}^{\cal N}\theta_ix_i\right)
   \langle\Ups_1(z_1,x_1;\theta_1)\dots
   \Ups_{\cal N}(z_{\cal N},x_{\cal N};\theta_{\cal N})\rangle
\label{affwardx}
\eea
Here we have introduced the differential operator
\be
 {\cal J}^{\cal N}\ =\ \sum_{i=1}^{\cal N}
  \left(x_i^2\pa_{x_i}-2j_ix_i\right)
\label{calJ}
\ee
It is noted that the middle identities in (\ref{confwardx}) 
and (\ref{affwardx}) follow from the first and third identities.
This is a simple consequence of $[L_1,L_{-1}]=2L_0$ 
and $[J_+,J_-]=2J_0$, respectively.
The first conformal and $SL(2,\mathbb{R})$ Ward identities merely impose
translation invariance on the correlators, allowing
us to express them solely in terms of differences, 
$z_i-z_j$ and $x_i-x_j$, between coordinates of the
same type.

The two sets of identities are very similar in nature, as the act of replacing
$(x_i,j_i,\theta_i)$ by $(z_i,-\D_i,-\mu_i)$ in the operators
appearing in (\ref{affwardx}) leads to (\ref{confwardx}).
Also, one of the sets of operators does not depend on 
the group coordinates $x_i$, while the other set of
operators does not depend on the conformal
coordinates $z_i$. 
We know that the form of two- and three-point
conformal blocks is fixed by the conformal Ward identities, 
cf. Section \ref{LogCFT}. We also know 
that the two- and three-point chiral blocks involving only 
generating-function primary fields factorize into
a `conformal` part and a `group' part.
It is therefore natural to expect that the general
two- and three-point chiral blocks may factor into a 
conformal part and a group part. Our analysis will support
this assertion. It is demonstrated in the process, though, that
one would miss a wealth of solutions by making such
a factorization ansatz too naively. Furthermore, a factorization
is not guaranteed a priori, so we will base our
analysis on very general ans\"atze for the form
of the two- and three-point chiral blocks.

\subsection{Two-point chiral blocks}

Let us first comment on the factorization ansatz alluded to above.
The naive approach would be to 
consider two-point chiral blocks constructed
by multiplying expressions of the form (\ref{2uni})
with similar expressions for the group part. To illustrate
the shortage of this procedure, we focus on the two-point chiral block of two
generating-function unified cells:
\bea
  \langle\Ups_1(z_1,x_1;\theta_1)\Ups_2(z_2,x_2;\theta_2)\rangle
 &=&\delta_{\D_1,\D_2}\frac{A^1
  \left(\mu_1+\mu_2-2\mu_1\mu_2\ln z_{12}\right)
  +A^{12}\mu_1\mu_2}{z_{12}^{\D_1+\D_2}}\nn
  &\times&\delta_{j_1,j_2}\{B^1
  \left(\theta_1+\theta_2+2\theta_1\theta_2\ln x_{12}\right)
  -B^{12}\theta_1\theta_2\}x_{12}^{j_1+j_2}\nn
  &=&\delta_{j_1,j_2}\frac{2A^1B^1(j_1+j_2+1)}{k+2}
   \theta_1\theta_2\frac{x_{12}^{j_1+j_2}}{z_{12}^{\D_1+\D_2}}
\label{2fact}
\eea
This could have been the end of the story in which case 
the only non-vanishing two-point chiral block of the individual
fields would have been the one containing two 
generating-function logarithmic
fields. Furthermore, no logarithmic singularity would occur.
As we will discuss presently, there are more non-trivial
solutions than this one.

We will base our analysis on the following ansatz
\be
 \langle\Ups_1(z_1,x_1;\theta_1)\Ups_2(z_2,x_2;\theta_2)\rangle
  \ =\ \{A(\theta_1,\theta_2)+B(\theta_1,\theta_2)\ln z_{12}
    +C(\theta_1,\theta_2)\ln x_{12}\}
    \frac{x_{12}^{2s}}{z_{12}^{2h}}
\label{2ansx}
\ee
where 
\be
 A(\theta_1,\theta_2)\ =\ A^0+A^1\theta_1+A^2\theta_2
  +A^{12}\theta_1\theta_2
\label{Ax}
\ee
and similarly for $B(\theta_1,\theta_2)$ and $C(\theta_1,\theta_2)$.
Correlators involving non-cellular primary fields correspond
to setting the associated $\theta$s equal to zero.
Due to the translational invariance (in both sets of coordinates)
of the ansatz, it suffices to impose the two identities involving
${\cal L}^{\cal N}$ and ${\cal J}^{\cal N}$. 
The third conformal Ward
identity (\ref{confwardx}) thus leads to the conditions
\bea
 0&=&\left(-h+\D_1+\frac{2j_1+1}{k+2}\theta_1\right)A(\theta_1,\theta_2)
  +\hf B(\theta_1,\theta_2)\nn
 0&=&\left(-h+\D_2+\frac{2j_2+1}{k+2}\theta_2\right)A(\theta_1,\theta_2)
  +\hf B(\theta_1,\theta_2)\nn
 0&=&\left(-h+\D_1+\frac{2j_1+1}{k+2}\theta_1\right)B(\theta_1,\theta_2)
  \ =\ \left(-h+\D_2+\frac{2j_2+1}{k+2}\theta_2\right)B(\theta_1,\theta_2)\nn
 0&=&\left(-h+\D_1+\frac{2j_1+1}{k+2}\theta_1\right)C(\theta_1,\theta_2)
  \ =\ \left(-h+\D_2+\frac{2j_2+1}{k+2}\theta_2\right)C(\theta_1,\theta_2)
\label{2cw}
\eea
whereas the third $SL(2,\mathbb{R})$ Ward identity corresponds to the conditions
\bea
 0&=&\left(s-j_1-\theta_1\right)A(\theta_1,\theta_2)
  +\hf C(\theta_1,\theta_2)
  \ =\ \left(s-j_2-\theta_2\right)A(\theta_1,\theta_2)
  +\hf C(\theta_1,\theta_2)\nn
 0&=&\left(s-j_1-\theta_1\right)B(\theta_1,\theta_2)
  \ =\ \left(s-j_2-\theta_2\right)B(\theta_1,\theta_2)\nn
 0&=&\left(s-j_1-\theta_1\right)C(\theta_1,\theta_2)
  \ =\ \left(s-j_2-\theta_2\right)C(\theta_1,\theta_2)
\label{2aw} 
\eea
We defer the analysis of these conditions to Appendix A.
It is noted, though, that as in the case of (non-affine) conformal Jordan 
cells \cite{logWard},
one may lose solutions for correlators involving non-cellular
fields if one simply sets the corresponding $\theta$s equal
to zero in the solution for unified cells only. Instead, examining the
conditions case by case (distinguished by the number
of generating-function unified cells appearing in the
generating-function two-point chiral block) as done in Appendix A
results in
\bea
 &&\langle\Ups_1(z_1,x_1;0)\Ups_2(z_2,x_2;0)\rangle
 \ =\ A^0W_2\nn
 &&\langle\Ups_1(z_1,x_1;\theta_1)\Ups_2(z_2,x_2;0)\rangle
 \ =\ A^1\theta_1W_2\nn
 &&\langle\Ups_1(z_1,x_1;\theta_1)\Ups_2(z_2,x_2;\theta_2)\rangle\nn
 &&\ \ \ \ \ \ \ \ =\ \left\{A^1\theta_1+A^1\theta_2+A^{12}\theta_1\theta_2
 -2A^1\theta_1\theta_2\left(\frac{j_1+j_2+1}{k+2}\ln z_{12}
   -\ln x_{12}\right)\right\}W_2
\label{2x}
\eea
Here we have introduced the abbreviation
\be
 W_2\ =\ \delta_{j_1,j_2}
  \frac{x_{12}^{j_1+j_2}}{z_{12}^{\D_1+\D_2}}
\label{W2}
\ee
and used that the identity
$A^2=A^1$ is required in the last two-point chiral block.
It is recalled that the weights are related to the spins
according to (\ref{Dj}).
In terms of the individual correlators, it follows from (\ref{2x}) that
\bea
 \langle\Phi_1(z_1,x_1)\Ups_2(z_2,x_2;0)\rangle
 &=&0\nn
 \langle\Psi_1(z_1,x_1)\Ups_2(z_2,x_2;0)\rangle
  &=&A^1W_2
\label{2x1i}
\eea
and
\bea
 \langle\Phi_1(z_1,x_1)\Phi_2(z_2,x_2)\rangle
  &=&0\nn
 \langle\Psi_1(z_1,x_1)\Phi_2(z_2,x_2)\rangle
  &=&A^1W_2\nn
 \langle\Psi_1(z_1,x_1)\Psi_2(z_2,x_2)\rangle
  &=&\left\{A^{12}-2A^1\left(
   \frac{j_1+j_2+1}{k+2}\ln z_{12}-\ln x_{12}\right)\right\}W_2
\label{2x2i}
\eea
The remaining two-point chiral blocks are obtained by appropriately 
permuting the indices. It is stressed that the structure
constant $A^1$ appearing in (\ref{2x1i}) a priori is independent of
the structure constant $A^1$ appearing in (\ref{2x2i}).

For a translational-invariant two-point chiral block, 
there is only one independent KZ equation. Referring to the ansatz
(\ref{2ansx}) or to the (loop-)projectively invariant
expressions (\ref{2x}), it may be written
\bea
 0&=&\left((k+2)\pa_{z_{12}}+\frac{-x_{12}^2\pa_{x_{12}}^2+
   2(j_1+j_2+\theta_1+\theta_2)x_{12}\pa_{x_{12}}
   -2(j_1+\theta_1)(j_2+\theta_2)}{z_{12}}\right)\nn
 &\times&\langle\Ups_1(z_1,x_1;\theta_1)\Ups_2(z_2,x_2;\theta_2)\rangle
\label{KZ2th}
\eea
Here one or both of the nilpotent parameters may vanish 
depending on the number of generating-function unified cells
there are in the correlator. 
It is straightforward to verify that the
generating-function two-point chiral blocks (\ref{2x}) satisfy this
KZ equation.

\subsection{Three-point chiral blocks}

Here we base our analysis on the ansatz
\bea
 &&\langle\Ups_1(z_1,x_1;\theta_1)\Ups_2(z_2,x_2;\theta_2)
  \Ups_3(z_3,x_3;\theta_3)\rangle\nn
 &=&
  \{A(\theta_1,\theta_2,\theta_3)
   +B_{12}(\theta_1,\theta_2,\theta_3)\ln z_{12}
   +B_{23}(\theta_1,\theta_2,\theta_3)\ln z_{23}
   +B_{13}(\theta_1,\theta_2,\theta_3)\ln z_{13}\nn
  &&+\ C_{12}(\theta_1,\theta_2,\theta_3)\ln x_{12}
   +C_{23}(\theta_1,\theta_2,\theta_3)\ln x_{23}
   +C_{13}(\theta_1,\theta_2,\theta_3)\ln x_{13}\nn
 &&+\ D_{11}(\theta_1,\theta_2,\theta_3)\ln^2 z_{12}
  +D_{12}(\theta_1,\theta_2,\theta_3)\ln z_{12}\ln z_{23}
  +D_{13}(\theta_1,\theta_2,\theta_3)\ln z_{12}\ln z_{13}\nn
  &&+\ D_{22}(\theta_1,\theta_2,\theta_3)\ln^2 z_{23}
  +D_{23}(\theta_1,\theta_2,\theta_3)\ln z_{23}\ln z_{13}
  +D_{33}(\theta_1,\theta_2,\theta_3)\ln^2 z_{13}\nn
 &&+\ E_{11}(\theta_1,\theta_2,\theta_3)\ln z_{12}\ln x_{12}
  +E_{12}(\theta_1,\theta_2,\theta_3)\ln z_{12}\ln x_{23}
  +E_{13}(\theta_1,\theta_2,\theta_3)\ln z_{12}\ln x_{13}\nn
  &&+\ E_{21}(\theta_1,\theta_2,\theta_3)\ln z_{23}\ln x_{12}
  +E_{22}(\theta_1,\theta_2,\theta_3)\ln z_{23}\ln x_{23}
  +E_{23}(\theta_1,\theta_2,\theta_3)\ln z_{23}\ln x_{13}\nn
 &&+\ E_{31}(\theta_1,\theta_2,\theta_3)\ln z_{13}\ln x_{12}
  +E_{32}(\theta_1,\theta_2,\theta_3)\ln z_{13}\ln x_{23}
  +E_{33}(\theta_1,\theta_2,\theta_3)\ln z_{13}\ln x_{13}\nn
  &&+\ F_{11}(\theta_1,\theta_2,\theta_3)\ln^2 x_{12}
  +F_{12}(\theta_1,\theta_2,\theta_3)\ln x_{12}\ln x_{23}
  +F_{13}(\theta_1,\theta_2,\theta_3)\ln x_{12}\ln x_{13}\nn
  &&+\ F_{22}(\theta_1,\theta_2,\theta_3)\ln^2 x_{23}
  +F_{23}(\theta_1,\theta_2,\theta_3)\ln x_{23}\ln x_{13}
  +F_{33}(\theta_1,\theta_2,\theta_3)\ln^2 x_{13}\}
 \frac{x_{12}^{s_{1}}x_{23}^{s_{2}}x_{13}^{s_{3}}}{
  z_{12}^{h_{1}}z_{23}^{h_{2}}z_{13}^{h_{3}}}
\label{3ansx}
\eea
where
\be
 A(\theta_1,\theta_2,\theta_3)\ =\ 
  A^0+A^1\theta_1+A^2\theta_2+A^3\theta_3
  +A^{12}\theta_1\theta_2+A^{23}\theta_2\theta_3+A^{13}\theta_1\theta_3
  +A^{123}\theta_1\theta_2\theta_3
\label{Athx}
\ee
and similarly for the other $\theta$-dependent structure 
constants: $B_{ij}(\theta_1,\theta_2,\theta_3)$, 
$C_{ij}(\theta_1,\theta_2,\theta_3)$, 
$D_{ij}(\theta_1,\theta_2,\theta_3)$,
$E_{ij}(\theta_1,\theta_2,\theta_3)$, and
$F_{ij}(\theta_1,\theta_2,\theta_3)$.
The conditions following from imposing the conformal and $SL(2,\mathbb{R})$
Ward identities are discussed in Appendix A.
This analysis leads to 
the following generating-function three-point chiral blocks:
\bea
 &&\langle\Ups_1(z_1,x_1;0)\Ups_2(z_2,x_2;0)\Ups_3(z_3,x_3;0)
  \rangle\ =\ A^0W_3\nn
 &&\langle\Ups_1(z_1,x_1;\theta_1)\Ups_2(z_2,x_2;0)\Ups_3(z_3,x_3;0)
  \rangle\nn
 &&\ \ =\ \left\{A^0+A^1\theta_1
  -A^0\theta_1\left(\frac{2j_1+1}{k+2}\ln\frac{z_{12}z_{13}}{z_{23}}
   -\ln\frac{x_{12}x_{13}}{x_{23}}\right)\right\}W_3\nn
 &&\langle\Ups_1(z_1,x_1;\theta_1)\Ups_2(z_2,x_2;\theta_2)
  \Ups_3(z_3,x_3;0)\rangle\nn
 &&\ \ =\ \left\{A^0+A^1\theta_1
  -A^0\theta_1\left(\frac{2j_1+1}{k+2}\ln\frac{z_{12}z_{13}}{z_{23}}
   -\ln\frac{x_{12}x_{13}}{x_{23}}\right)\right.\nn
 &&\ \ \ \ +\ A^2\theta_2
   -A^0\theta_2\left(\frac{2j_2+1}{k+2}\ln\frac{z_{12}z_{23}}{z_{13}}
   -\ln\frac{x_{12}x_{23}}{x_{13}}\right)\nn
 &&\ \ \ \ +\ A^{12}\theta_1\theta_2-A^1\theta_1\theta_2\left(\frac{2j_2+1}{k+2}
  \ln\frac{z_{12}z_{23}}{z_{13}}-\ln\frac{x_{12}x_{23}}{x_{13}}\right)
   -A^2\theta_1\theta_2\left(\frac{2j_1+1}{k+2}\ln\frac{z_{12}z_{13}}{z_{23}}
   -\ln\frac{x_{12}x_{13}}{x_{23}}\right)\nn
 &&\ \ \ \ +\left. A^0\theta_1\theta_2
   \left(\frac{2j_1+1}{k+2}\ln\frac{z_{12}z_{13}}{z_{23}}
   -\ln\frac{x_{12}x_{13}}{x_{23}}\right)\left(\frac{2j_2+1}{k+2}
  \ln\frac{z_{12}z_{23}}{z_{13}}-\ln\frac{x_{12}x_{23}}{x_{13}}\right)
    \right\}W_3\nn
 &&\langle\Ups_1(z_1,x_1;\theta_1)\Ups_2(z_2,x_2;\theta_2)
    \Ups_3(z_3,x_3;\theta_3)\rangle\nn 
 &&\ \ =\ \left\{A^1\theta_1+A^2\theta_2+A^3\theta_3+A^{12}\theta_1\theta_2
  -A^1\theta_1\theta_2\left(\frac{2j_2+1}{k+2}\ln\frac{z_{12}z_{23}}{z_{13}}
   -\ln\frac{x_{12}x_{23}}{x_{13}}\right)\right.\nn
 &&\ \ \ \ -\ A^2\theta_1\theta_2\left(\frac{2j_1+1}{k+2}\ln\frac{z_{12}z_{13}}{z_{23}}
      -\ln\frac{x_{12}x_{13}}{x_{23}}\right)\nn
 &&\ \ \ \ +\ A^{23}\theta_2\theta_3
    -A^2\theta_2\theta_3\left(\frac{2j_3+1}{k+2}\ln\frac{z_{23}z_{13}}{z_{12}}
   -\ln\frac{x_{23}x_{13}}{x_{12}}\right)
   -A^3\theta_2\theta_3\left(\frac{2j_2+1}{k+2}\ln\frac{z_{12}z_{23}}{z_{13}}
      -\ln\frac{x_{12}x_{23}}{x_{13}}\right)\nn
 &&\ \ \ \ +\ A^{13}\theta_1\theta_3
    -A^3\theta_1\theta_3\left(\frac{2j_1+1}{k+2}\ln\frac{z_{12}z_{13}}{z_{23}}
   -\ln\frac{x_{12}x_{13}}{x_{23}}\right)
   -A^1\theta_1\theta_3\left(\frac{2j_3+1}{k+2}\ln\frac{z_{23}z_{13}}{z_{12}}
      -\ln\frac{x_{23}x_{13}}{x_{12}}\right)\nn
 &&\ \ \ \ +\ A^{123}\theta_1\theta_2\theta_3
   -A^{12}\theta_1\theta_2\theta_3\left(\frac{2j_3+1}{k+2}
     \ln\frac{z_{23}z_{13}}{z_{12}}-\ln\frac{x_{23}x_{13}}{x_{12}}\right)
    \nn
&&\ \ \ \ -\ A^{23}\theta_1\theta_2\theta_3\left(\frac{2j_1+1}{k+2}
     \ln\frac{z_{12}z_{13}}{z_{23}}-\ln\frac{x_{12}x_{13}}{x_{23}}\right)
   -A^{13}\theta_1\theta_2\theta_3\left(\frac{2j_2+1}{k+2}
     \ln\frac{z_{12}z_{23}}{z_{13}}-\ln\frac{x_{12}x_{23}}{x_{13}}\right)\nn
 &&\ \ \ \ +\ A^1\theta_1\theta_2\theta_3
   \left(\frac{2j_2+1}{k+2}\ln\frac{z_{12}z_{23}}{z_{13}}
   -\ln\frac{x_{12}x_{23}}{x_{13}}\right)\left(\frac{2j_3+1}{k+2}
   \ln\frac{z_{23}z_{13}}{z_{12}}-\ln\frac{x_{23}x_{13}}{x_{12}}\right)\nn
&&\ \ \ \ +\ A^2\theta_1\theta_2\theta_3
   \left(\frac{2j_1+1}{k+2}\ln\frac{z_{12}z_{13}}{z_{23}}
   -\ln\frac{x_{12}x_{13}}{x_{23}}\right)\left(\frac{2j_3+1}{k+2}
   \ln\frac{z_{23}z_{13}}{z_{12}}-\ln\frac{x_{23}x_{13}}{x_{12}}\right)\nn
&&\ \ \ \ +\left. A^3\theta_1\theta_2\theta_3
   \left(\frac{2j_1+1}{k+2}\ln\frac{z_{12}z_{13}}{z_{23}}
   -\ln\frac{x_{12}x_{13}}{x_{23}}\right)\left(\frac{2j_2+1}{k+2}
   \ln\frac{z_{12}z_{23}}{z_{13}}-\ln\frac{x_{12}x_{23}}{x_{13}}\right) 
 \right\}W_3
\label{3x}
\eea
Here we have introduced the abbreviation
\be
 W_3\ =\ \frac{x_{12}^{j_1+j_2-j_3}x_{23}^{-j_1+j_2+j_3}x_{13}^{j_1-j_2+j_3}}{
   z_{12}^{\D_1+\D_2-\D_3}z_{23}^{-\D_1+\D_2+\D_3}z_{13}^{\D_1-\D_2+\D_3}}
\label{W3}
\ee
In terms of individual correlators (besides the one for non-cellular
primary fields only, which has been already listed in (\ref{3x})), 
we thus have
\bea
 \langle\Phi_1(z_1,x_1)\Ups_2(z_2,x_2;0)\Ups_3(z_3,x_3;0)
  \rangle&=&A^0W_3\nn
 \langle\Psi_1(z_1,x_1)\Ups_2(z_2,x_2;0)\Ups_3(z_3,x_3;0)
  \rangle&=&\left\{A^1
  -A^0\left(\frac{2j_1+1}{k+2}\ln\frac{z_{12}z_{13}}{z_{23}}
   -\ln\frac{x_{12}x_{13}}{x_{23}}\right)\right\}W_3
\label{3x1i}
\eea
and
\bea
 &&\langle\Phi_1(z_1,x_1)\Phi_2(z_2,x_2)\Ups_3(z_3,x_3;0)
  \rangle\ =\ A^0W_3\nn
 &&\langle\Psi_1(z_1,x_1)\Phi_2(z_2,x_2)\Ups_3(z_3,x_3;0)
  \rangle\ =\ \left\{A^1
  -A^0\left(\frac{2j_1+1}{k+2}\ln\frac{z_{12}z_{13}}{z_{23}}
   -\ln\frac{x_{12}x_{13}}{x_{23}}\right)\right\}W_3\nn
 &&\langle\Psi_1(z_1,x_1)\Psi_2(z_2,x_2)\Ups_3(z_3,x_3;0)
  \rangle\nn
 &&\ \ =\ \left\{A^{12}-A^1\left(\frac{2j_2+1}{k+2}
  \ln\frac{z_{12}z_{23}}{z_{13}}-\ln\frac{x_{12}x_{23}}{x_{13}}\right)
   -A^2\left(\frac{2j_1+1}{k+2}\ln\frac{z_{12}z_{13}}{z_{23}}
   -\ln\frac{x_{12}x_{13}}{x_{23}}\right)\right.\nn
 &&\ \ \ \ \ \ \ +\left. A^0\left(\frac{2j_1+1}{k+2}\ln\frac{z_{12}z_{13}}{z_{23}}
   -\ln\frac{x_{12}x_{13}}{x_{23}}\right)\left(\frac{2j_2+1}{k+2}
   \ln\frac{z_{12}z_{23}}{z_{13}}-\ln\frac{x_{12}x_{23}}{x_{13}}\right)
  \right\}W_3
\label{3x2i}
\eea
and
\bea
 &&\langle\Phi_1(z_1,x_1)\Phi_2(z_2,x_2)\Phi_3(z_3,x_3)
  \rangle\ =\ 0\nn
 &&\langle\Psi_1(z_1,x_1)\Phi_2(z_2,x_2)\Phi_3(z_3,x_3)
  \rangle\ =\ A^1W_3\nn
 &&\langle\Psi_1(z_1,x_1)\Psi_2(z_2,x_2)\Phi_3(z_3,x_3)
  \rangle\nn
 &&\ \ \ \ =\ \left\{A^{12}
   -A^1\left(\frac{2j_2+1}{k+2}\ln\frac{z_{12}z_{23}}{z_{13}}
   -\ln\frac{x_{12}x_{23}}{x_{13}}\right)
   -A^2\left(\frac{2j_1+1}{k+2}\ln\frac{z_{12}z_{13}}{z_{23}}
      -\ln\frac{x_{12}x_{13}}{x_{23}}\right)  \right\}W_3\nn
 &&\langle\Psi_1(z_1,x_1)\Psi_2(z_2,x_2)\Psi_3(z_3,x_3)
  \rangle\nn
 &&\ \ \ \ =\ \left\{A^{123}
   -A^{12}\left(\frac{2j_3+1}{k+2}
     \ln\frac{z_{23}z_{13}}{z_{12}}-\ln\frac{x_{23}x_{13}}{x_{12}}\right)
    -A^{23}\left(\frac{2j_1+1}{k+2}
     \ln\frac{z_{12}z_{13}}{z_{23}}-\ln\frac{x_{12}x_{13}}{x_{23}}\right)
   \right.\nn
 &&\ \ \ \ \ \ \ \ -\ A^{13}\left(\frac{2j_2+1}{k+2}
     \ln\frac{z_{12}z_{23}}{z_{13}}-\ln\frac{x_{12}x_{23}}{x_{13}}\right)\nn
 &&\ \ \ \ \ \ \ \ +\ A^1\left(\frac{2j_2+1}{k+2}\ln\frac{z_{12}z_{23}}{z_{13}}
   -\ln\frac{x_{12}x_{23}}{x_{13}}\right)\left(\frac{2j_3+1}{k+2}
   \ln\frac{z_{23}z_{13}}{z_{12}}-\ln\frac{x_{23}x_{13}}{x_{12}}\right)\nn
&&\ \ \ \ \ \ \ \ +\ A^2\left(\frac{2j_1+1}{k+2}\ln\frac{z_{12}z_{13}}{z_{23}}
   -\ln\frac{x_{12}x_{13}}{x_{23}}\right)\left(\frac{2j_3+1}{k+2}
   \ln\frac{z_{23}z_{13}}{z_{12}}-\ln\frac{x_{23}x_{13}}{x_{12}}\right)\nn
&&\ \ \ \ \ \ \ \ +\left. A^3\left(\frac{2j_1+1}{k+2}\ln\frac{z_{12}z_{13}}{z_{23}}
   -\ln\frac{x_{12}x_{13}}{x_{23}}\right)\left(\frac{2j_2+1}{k+2}
   \ln\frac{z_{12}z_{23}}{z_{13}}-\ln\frac{x_{12}x_{23}}{x_{13}}\right) 
 \right\}W_3
\label{3x3i}
\eea
The remaining three-point chiral blocks are obtained by appropriately 
permuting the indices. 

For a translational-invariant three-point chiral block, 
there are two independent KZ equations, cf. (\ref{sumKZ}). 
Referring to the ansatz (\ref{3ansx}) or to the (loop-)projectively invariant
expressions (\ref{3x}), they may be written
\be
 0\ =\ KZ_i\langle\Ups_1(z_1,x_1;\theta_1)\Ups_2(z_2,x_2;\theta_2)
  \Ups_3(z_3,x_3;\theta_3)\rangle,\ \ \ \ \ \ \ i=1,2
\label{KZ3th0}
\ee
where 
\bea
 KZ_1&=&(k+2)\pa_{z_1}-\frac{2D_0(x_1,\theta_1)D_0(x_2,\theta_2)
  +D_+(x_1,\theta_1)D_-(x_2,\theta_2)
  +D_-(x_1,\theta_1)D_+(x_2,\theta_2)}{z_{12}}\nn
   &&\ \ \ \ \ \ -\ \frac{2D_0(x_1,\theta_1)D_0(x_3,\theta_3)
  +D_+(x_1,\theta_1)D_-(x_3,\theta_3)
  +D_-(x_1,\theta_1)D_+(x_3,\theta_3)}{z_{13}}\nn
 KZ_2&=&(k+2)\pa_{z_2}+\frac{2D_0(x_1,\theta_1)D_0(x_2,\theta_2)
  +D_+(x_1,\theta_1)D_-(x_2,\theta_2)
  +D_-(x_1,\theta_1)D_+(x_2,\theta_2)}{z_{12}}\nn
   &&\ \ \ \ \ \ -\ \frac{2D_0(x_2,\theta_2)D_0(x_3,\theta_3)
  +D_+(x_2,\theta_2)D_-(x_3,\theta_3)
  +D_-(x_2,\theta_2)D_+(x_3,\theta_3)}{z_{23}}
\label{KZ3th}
\eea
In these expressions, one, two or all three 
of the nilpotent parameters may vanish 
depending on the number of generating-function unified cells
there are in the correlator. 
It is straightforward, though rather tedious, to verify that the
generating-function three-point chiral blocks (\ref{3x}) satisfy these
KZ equations.

\subsection{In terms of spins with nilpotent parts}

Here we wish to extend to the logarithmic WZW model
the idea put forward in \cite{MRS} that correlators in
logarithmic CFT may be represented compactly by considering
conformal weights with nilpotent parts $\Dh+\thh$. 
The most general results of this
kind for two- and three-point conformal blocks
were found in \cite{logWard} and are given
above as (\ref{2unith}) and (\ref{3th}).
As already indicated, we will here associate the generalized
spin $j_i+\theta_i$ to the generating-function
unified cell $\Ups_i(z_i,x_i;\theta_i)$. The corresponding generalized 
conformal weight thus reads $\D_i+\mu_i$ where 
$\mu_i=(2j_i+1)\theta_i/(k+2)$. This allows us to express
the generating-function correlators for two- and three-point chiral blocks 
in the following simple way:
\bea
 \langle\Ups_1(z_1,x_1;0)\Ups_2(z_2,x_2;0)\rangle
 &=&\delta_{j_1,j_2}A^0\frac{x_{12}^{j_1+j_2}}{z_{12}^{\D_1+\D_2}}\nn
 \langle\Ups_1(z_1,x_1;\theta_1)\Ups_2(z_2,x_2;0)\rangle
 &=&\delta_{j_1,j_2}A^1\theta_1\frac{x_{12}^{(j_1+\theta_1)+j_2}}{
  z_{12}^{(\D_1+\mu_1)+\D_2}}\nn
 \langle\Ups_1(z_1,x_1;\theta_1)\Ups_2(z_2,x_2;\theta_2)\rangle
 &=&\delta_{j_1,j_2}\left\{A^1\theta_1+A^1\theta_2+A^{12}\theta_1\theta_2\right\}
    \frac{x_{12}^{(j_1+\theta_1)+(j_2+\theta_2)}}{z_{12}^{(\D_1+\mu_1)+(\D_2+\mu_2)}}
\label{2xth}
\eea
and
\bea
 && \langle\Ups_1(z_1,x_1;0)\Ups_2(z_2,x_2;0)\Ups_3(z_3,x_3;0)
  \rangle
  \ =\ A^0\frac{x_{12}^{j_1+j_2-j_3}x_{23}^{-j_1+j_2+j_3}x_{13}^{j_1-j_2+j_3}}{
   z_{12}^{\D_1+\D_2-\D_3}z_{23}^{-\D_1+\D_2+\D_3}z_{13}^{\D_1-\D_2+\D_3}}\nn
 &&\langle\Ups_1(z_1,x_1;\theta_1)\Ups_2(z_2,x_2;0)\Ups_3(z_3,x_3;0)
  \rangle\ =\ \left\{A^0+A^1\theta_1\right\}\nn
 &&\ \ \ \ \times\ \frac{x_{12}^{(j_1+\theta_1)+j_2-j_3}
   x_{23}^{-(j_1+\theta_1)+j_2+j_3}x_{13}^{(j_1+\theta_1)-j_2+j_3}}{
   z_{12}^{(\D_1+\mu_1)+\D_2-\D_3}z_{23}^{-(\D_1+\mu_1)+\D_2+\D_3}
   z_{13}^{(\D_1+\mu_1)-\D_2+\D_3}}\nn
  &&\langle\Ups_1(z_1,x_1;\theta_1)\Ups_2(z_2,x_2;\theta_2)
  \Ups_3(z_3,x_3;0)\rangle
   \ =\ \left\{A^0+A^1\theta_1+A^2\theta_2+A^{12}\theta_1\theta_2\right\}\nn 
  &&\ \ \ \ \times\ 
   \frac{x_{12}^{(j_1+\theta_1)+(j_2+\theta_2)-j_3}
   x_{23}^{-(j_1+\theta_1)+(j_2+\theta_2)+j_3}
   x_{13}^{(j_1+\theta_1)-(j_2+\theta_2)+j_3}}{
   z_{12}^{(\D_1+\mu_1)+(\D_2+\mu_2)-\D_3}
   z_{23}^{-(\D_1+\mu_1)+(\D_2+\mu_2)+\D_3}
   z_{13}^{(\D_1+\mu_1)-(\D_2+\mu_2)+\D_3}} \nn
 &&\langle\Ups_1(z_1,x_1;\theta_1)\Ups_2(z_2,x_2;\theta_2)
    \Ups_3(z_3,x_3;\theta_3)\rangle\nn 
 &&\ \ \ \ =\ \left\{A^1\theta_1+A^2\theta_2+A^3\theta_3
   +A^{12}\theta_1\theta_2
   +A^{23}\theta_2\theta_3+A^{13}\theta_1\theta_3
   +A^{123}\theta_1\theta_2\theta_3\right\}\nn
 &&\ \ \ \ \times\ \frac{x_{12}^{(j_1+\theta_1)+(j_2+\theta_2)-(j_3+\theta_3)}
  x_{23}^{-(j_1+\theta_1)+(j_2+\theta_2)+(j_3+\theta_3)}
  x_{13}^{(j_1+\theta_1)-(j_2+\theta_2)+(j_3+\theta_3)}}{
   z_{12}^{(\D_1+\mu_1)+(\D_2+\mu_2)-(\D_3+\mu_3)}
   z_{23}^{-(\D_1+\mu_1)+(\D_2+\mu_2)+(\D_3+\mu_3)}
   z_{13}^{(\D_1+\mu_1)-(\D_2+\mu_2)+(\D_3+\mu_3)}} 
\label{3xth}
\eea
The remaining combination in (\ref{2xth}) and the remaining 
four combinations in (\ref{3xth}) are obtained by appropriate
permutations in the indices.
Thus confirming our assertion, 
it follows that the two- and three-point chiral blocks factor 
into a conformal part and an $SL(2,\mathbb{R})$ group part.
The extra degrees of freedom in (\ref{2xth}) and (\ref{3xth}) compared
to the incomplete result (\ref{2fact}) are
contained in the fact that the $\theta$-dependent structure constants
in (\ref{2xth}) and (\ref{3xth}) do not necessarily factor as in
(\ref{2fact}).
It is noted that the present factorization into a conformal part and an 
$SL(2,\mathbb{R})$ group part is not evident a priori, while our analysis has
demonstrated its validity.
These compact representations constitute a significant simplification
of the results given above (and derived in Appendix A).
The verification of the KZ equations is particularly simple when
the correlators are expressed in this way.

\subsection{Hierarchical structures for chiral blocks}

Based on ideas discussed in \cite{FloOPE,RAK}, it was found in \cite{logWard}
that the conformal blocks involving logarithmic fields may be represented
in terms of derivatives with respect to the conformal weights.
This was reviewed in Section \ref{derive}. Here we wish to extend this idea
to the two- and three-point chiral blocks of the logarithmic WZW model
introduced above. It is found that hierarchical structures similar to
the ones discussed in Section \ref{derive} apply in the affine case.

First, it is observed that
acting on either $W_2$ or $W_3$, 
we may substitute derivatives with respect to the spins
by multiplicative factors according to
\be
 \pa_{j_1}\ =\ \pa_{j_2}\ \rightarrow\ 
  -2\frac{j_1+j_2+1}{k+2}\ln z_{12}+2\ln x_{12}
\label{Dln2}
\ee
or 
\bea
 \pa_{j_1}&\rightarrow&-\frac{2j_1+1}{k+2}\ln\frac{z_{12}{z_{13}}}{z_{23}}
   +\ln\frac{x_{12}{x_{13}}}{x_{23}}\nn
 \pa_{j_2}&\rightarrow&-\frac{2j_2+1}{k+2}\ln\frac{z_{12}{z_{23}}}{z_{13}}
   +\ln\frac{x_{12}{x_{23}}}{x_{13}}\nn 
 \pa_{j_3}&\rightarrow&-\frac{2j_3+1}{k+2}\ln\frac{z_{23}{z_{13}}}{z_{12}}
   +\ln\frac{x_{23}{x_{13}}}{x_{12}}
\label{Dln3}
\eea
respectively. This simple observation allows us to represent
the correlators involving logarithmic fields as follows:
\bea
  \langle\Psi_1(z_1,x_1)\Ups_2(z_2,x_2;0)\rangle&=&A^1W_2\nn
  \langle\Psi_1(z_1,x_1)\Phi_2(z_2,x_2)\rangle&=&A^1W_2\nn
  \langle\Psi_1(z_1,x_1)\Psi_2(z_2,x_2)\rangle&=&\left(A^{12}+A^2\pa_{j_1}
    +A^1\pa_{j_2}\right)W_2\nn
   \langle\Psi_1(z_1,x_1)\Ups_2(z_2,x_2;0)
   \Ups_3(z_3,x_3;0)\rangle&=&\left(A^1+A^0\pa_{j_1}\right)W_3\nn
 \langle\Psi_1(z_1,x_1)\Phi_2(z_2,x_2)
   \Ups_3(z_3,x_3;0)\rangle&=&\left(A^1+A^0\pa_{j_1}\right)W_3\nn
 \langle\Psi_1(z_1,x_2)\Psi_2(z_2,x_2)
   \Ups_3(z_3,x_3;0)\rangle&=&\left(A^{12}+A^1\pa_{j_2}
    +A^2\pa_{j_1}+A^0\pa_{j_1}\pa_{j_2}\right)W_3\nn
   \langle\Psi_1(z_1,x_1)\Phi_2(z_2,x_2)
   \Phi_3(z_3,x_3)\rangle&=&A^1W_3\nn
 \langle\Psi_1(z_1,x_1)\Psi_2(z_2,x_2)
   \Phi_3(z_3,x_3)\rangle&=&\left(A^{12}+A^2\pa_{j_1}
    +A^1\pa_{j_2}\right)W_3\nn
 \langle\Psi_1(z_1,x_1)\Psi_2(z_2,x_2)
   \Psi_3(z_3,x_3)\rangle&=&
  \left(A^{123}+A^{23}\pa_{j_1}+A^{13}\pa_{j_2}
   +A^{12}\pa_{j_3}\right.\nn
  &&+\left.A^3\pa_{j_1}\pa_{j_2}+A^1\pa_{j_2}\pa_{j_3}
   +A^2\pa_{j_1}\pa_{j_3}\right)W_3
\label{3Dx}
\eea 
in addition to expressions obtained by appropriately permuting
the indices.
One may therefore represent the correlators hierarchically as
\bea
 \langle\Psi_1(z_1,x_1)\Ups_2(z_2,x_2;0)\rangle&=&A^1W_2
   +\pa_{j_1}\langle\Phi_1(z_1,x_1)\Ups_2(z_2,x_2;0)\rangle\nn
  \langle\Psi_1(z_1,x_1)\Phi_2(z_2,x_2)\rangle&=&A^1W_2
    +\pa_{j_1}\langle\Phi_1(z_1,x_1)\Phi_2(z_2,x_2)\rangle\nn
  \langle\Psi_1(z_1,x_1)\Psi_2(z_2,x_2)\rangle&=&A^{12}W_2
   +\pa_{j_1}\langle\Phi_1(z_1,x_1)\Psi_2(z_2,x_2)\rangle
    +\pa_{j_2}\langle\Psi_1(z_1,x_1)\Phi_2(z_2,x_2)\rangle\nn
  &-&\pa_{j_1}\pa_{j_2}\langle\Phi_1(z_1,x_1)\Phi_2(z_2,x_2)\rangle
\label{2hx}
\eea
in the case of two-point chiral blocks, and
\bea
   \langle\Psi_1(z_1,x_1)\Ups_2(z_2,x_2;0)
   \Ups_3(z_3,x_3;0)\rangle&=&A^1W_3
   +\pa_{j_1}\langle\Phi_1(z_1,x_1)\Ups_2(z_2,x_2;0)
   \Ups_3(z_3,x_3;0)\rangle\nn
 \langle\Psi_1(z_1,x_1)\Phi_2(z_2,x_2)
   \Ups_3(z_3,x_3;0)\rangle&=&A^1W_3
   +\pa_{j_1}\langle\Phi_1(z_1,x_1)\Phi_2(z_2,x_2)
   \Ups_3(z_3,x_3;0)\rangle\nn
 \langle\Psi_1(z_1,x_1)\Psi_2(z_2,x_2)
   \Ups_3(z_3,x_3;0)\rangle&=&A^{12}W_3
    +\pa_{j_1}\langle\Phi_1(z_1,x_1)\Psi_2(z_2,x_2)
      \Ups_3(z_3,x_3;0)\rangle\nn
  &+&\pa_{j_2}\langle\Psi_1(z_1,x_1)\Phi_2(z_2,x_2)
      \Ups_3(z_3,x_3;0)\rangle\nn
  &-&\pa_{j_1}\pa_{j_2}\langle\Phi_1(z_1,x_1)\Phi_2(z_2,x_2)
    \Ups_3(z_3,x_3;0)\rangle\nn
   \langle\Psi_1(z_1,x_1)\Phi_2(z_2,x_2)
   \Phi_3(z_3,x_3)\rangle&=&A^1W_3
   +\pa_{j_1}\langle\Phi_1(z_1,x_1)\Phi_2(z_2,x_2)
   \Phi_3(z_3,x_3)\rangle\nn
 \langle\Psi_1(z_1,x_1)\Psi_2(z_2,x_2)
   \Phi_3(z_3,x_3)\rangle&=&A^{12}W_3
    +\pa_{j_1}\langle\Phi_1(z_1,x_1)\Psi_2(z_2,x_2)
      \Phi_3(z_3,x_3)\rangle\nn
 &+&\pa_{j_2}\langle\Psi_1(z_1,x_1)\Phi_2(z_2,x_2)
      \Phi_3(z_3,x_3)\rangle\nn
  &-&\pa_{j_1}\pa_{j_2}\langle\Phi_1(z_1,x_1)\Phi_2(z_2,x_2)
    \Phi_3(z_3,x_3)\rangle\nn
 \langle\Psi_1(z_1,x_1)\Psi_2(z_2,x_2)
   \Psi_3(z_3,x_3)\rangle&=&A^{123}W_3
   +\pa_{j_1}\langle\Phi_1(z_1,x_1)\Psi_2(z_2,x_2)
   \Psi_3(z_3,x_3)\rangle\nn
 &+&\pa_{j_2}\langle\Psi_1(z_1,x_1)\Phi_2(z_2,x_2)
   \Psi_3(z_3,x_3)\rangle\nn
 &+&\pa_{j_3}\langle\Psi_1(z_1,x_1)\Psi_2(z_2,x_2)
   \Phi_3(z_3,x_3)\rangle\nn
 &-&\pa_{j_1}\pa_{j_2}\langle\Phi_1(z_1,x_1)\Phi_2(z_2,x_2)
   \Psi_3(z_3,x_3)\rangle\nn
  &-&\pa_{j_2}\pa_{j_3}\langle\Psi_1(z_1,x_1)\Phi_2(z_2,x_2)
   \Phi_3(z_3,x_3)\rangle\nn
 &-&\pa_{j_1}\pa_{j_3}\langle\Phi_1(z_1,x_1)\Psi_2(z_2,x_2)
   \Phi_3(z_3,x_3)\rangle\nn
 &+&\pa_{j_1}\pa_{j_2}\pa_{j_3}\langle\Phi_1(z_1,x_1)\Phi_2(z_2,x_2)
   \Phi_3(z_3,x_3)\rangle
\label{3hx}
\eea
in the case of three-point chiral blocks. As above, the remaining correlators 
may be obtained by appropriately permuting the indices.

\section{Hamiltonian reduction}
\label{hamred}

It is well known that $SL(2,\mathbb{R})$ WZW models may be linked to conformal
minimal models via hamiltonian reduction \cite{Bel,Pol,BO,FF}.
A precise description of this reduction was given at the level
of correlators in \cite{FGPP,GP}, while a simple and direct proof
of this description 
was presented in \cite{PRYham,thesis} based on \cite{PRYblocks, thesis}.
The basic idea in this context is to start with an ${\cal N}$-point
chiral block of generating-function primary fields in the affine model, 
in which case the corresponding ${\cal N}$-point conformal block in the
CFT is obtained by setting $x_i=z_i$ for $i=1,\dots,{\cal N}$ \cite{FGPP,GP}.
This was refined a bit in \cite{PRYham,thesis} where it was discussed
how the procedure may be performed in two steps by first
setting $x_i=xz_i$ followed by fixing the common proportionality
constant to $x=1$, for example. 

Our current situation is quite
simple, though, since we are only interested in two- and three-point
functions and in their form rather than the relations
between structure constants and their dependencies on the spins
and conformal weights. The objective here is therefore to
study whether a naive extension of the hamiltonian-reduction
principle setting $x_i=z_i$ applies to the logarithmic correlators
found above. That is, we wish to show that the two- and three-point
chiral blocks (\ref{2x}) and (\ref{3x})
reduce to the two- and three-point conformal blocks 
(\ref{2uni}) and (\ref{3}), respectively, 
upon setting $x_i=z_i$. The conformal weights, $\Dh_i$, 
in the resulting logarithmic CFT should then be given by
\be
 \Dh_i\ =\ \D_i-j_i\ =\ \frac{j_i(j_i+1)}{k+2}-j_i
\label{Dhch}
\ee
whereas the central charges are related as
$\hat{c}=c-6k-2=\frac{3k}{k+2}-6k-2$.
It is emphasized that we are only concerned with the
form of the correlators, not the various dependencies
of the structure constants.  

The reductions are straightforward to analyze 
when the correlators are expressed compactly in terms of spins and
conformal weights with nilpotent parts. We thus wish to examine the
link between (\ref{2xth}) and (\ref{3xth}) on one hand and 
(\ref{2unith}) and (\ref{3th}) on the other hand.
It follows that the affine correlators reduce
to the conformal ones with conformal weights given in (\ref{Dhch}),
if the identifications $x_i=z_i$ for
$i=1,\dots,{\cal N}$ are accompanied by
\be
 \thh_i\ =\ \mu_i-\theta_i\ =\ \left(\frac{2j_i+1}{k+2}-1\right)
  \theta_i,\ \ \ \ \ \ \ i\ =\ 1,\dots,{\cal N}
\label{thhmu}
\ee
and (for $j_i,j_{i'}\neq(k+1)/2$) the renormalizations
\bea
 \Ah^0&=&A^0\nn
 \Ah^i&=&\frac{A^i}{\frac{2j_i+1}{k+2}-1},\ \ \ \ \ \ \ 1\leq i\leq3\nn
 \Ah^{ii'}&=&\frac{A^{ii'}}{\left(\frac{2j_i+1}{k+2}-1\right)
  \left(\frac{2j_{i'}+1}{k+2}-1\right)},\ \ \ \ \ \ \ 1\leq i<i'\leq3\nn
 \Ah^{123}&=&\frac{A^{123}}{\left(\frac{2j_1+1}{k+2}-1\right)
  \left(\frac{2j_2+1}{k+2}-1\right)\left(\frac{2j_3+1}{k+2}-1\right)}
\label{AhA}
\eea
The apparent subtlety in the case of two-point functions, composed 
of generating-function unified cells only, is resolved by the 
Kronecker delta function in $j_1$ and $j_2$ appearing in (\ref{2xth}).

In the exceptional case where $j_i=(k+1)/2$, this hamiltonian-reduction
procedure corresponds to formally replacing $\Ups_i(z_i,x_i;\theta_i)$
by the non-cellular primary field $\Ups(z_i;0)$, cf. (\ref{thhmu}), in which case
the renormalizations (\ref{AhA}) involving $j_i$ no longer apply.
The representation-theoretical mechanism underlying this reduction
in logarithmic nature remains to be understood.

Following \cite{FloCluster}, primary fields are called 
{\em proper primary} 
if their operator-product expansions
with each other cannot produce a logarithmic field.
It is argued in \cite{FloCluster} (see also \cite{GG}) that 
the structure constants of three-point conformal blocks 
not involving improper primary fields 
are related. According to \cite{logWard} and in
the notation used above, such conformal blocks
are obtained by setting
\be
 \Ah^1\ =\ \Ah^2\ =\ \Ah^3,\ \ \ \ \ \ \ \Ah^{12}\ =\ \Ah^{23}\ =\ \Ah^{13}
\label{clustercond}
\ee
This class of restricted three-point conformal blocks can be
reached by hamiltonian reduction of a particular subset of the three-point
chiral blocks in the affine case. This is quite obvious in the framework
with generalized spins and generalized conformal weights, cf. (\ref{AhA}).
One merely sets
\bea
 \frac{A^1}{\frac{2j_1+1}{k+2}-1}
 &=&\frac{A^2}{\frac{2j_2+1}{k+2}-1}
 \ =\ \frac{A^3}{\frac{2j_3+1}{k+2}-1}\nn
 \frac{A^{12}}{\left(\frac{2j_1+1}{k+2}-1\right)\left(\frac{2j_2+1}{k+2}-1\right)}
  &=& \frac{A^{23}}{\left(\frac{2j_2+1}{k+2}-1\right)\left(\frac{2j_3+1}{k+2}-1\right)}
 \ =\ \frac{A^{13}}{\left(\frac{2j_1+1}{k+2}-1\right)\left(\frac{2j_3+1}{k+2}-1\right)}
\label{Aclcond}
\eea
in the last chiral block in (\ref{3xth}). 
Hamiltonian reduction then reproduces the last three-point
conformal block in (\ref{3th}) with (\ref{clustercond}) satisfied.

\section{Conclusion}
\label{concl}

We have studied a particular type of logarithmic extension of $SL(2,\mathbb{R})$
WZW models. It is based on the introduction of affine Jordan cells
constructed as multiplets of quasi-primary fields organized in 
indecomposable representations of the Lie algebra $sl(2)$.
We have found the general solution to the simultaneously imposed set of
conformal and $SL(2,\mathbb{R})$ Ward identities
for two- and three-point chiral blocks. These correlators may involve
logarithmic terms and may be represented compactly by considering
spins with nilpotent parts. The chiral blocks have been found to
exhibit hierarchical structures obtained by computing derivatives
with respect to the spins.
A set of KZ equations, appropriately modified to cover affine Jordan cells,
have been derived, and the chiral blocks have been shown to satisfy
these equations.
It has been also demonstrated that
a simple and well-established prescription for hamiltonian reduction at the level
of correlators extends straightforwardly to the logarithmic 
correlators as the latter reduce to the known results for
two- and three-point conformal blocks in logarithmic CFT.

We find it natural to say that our results pertain to affine 
Jordan cells of rank two. This is supported in part by
the fact that hamiltonian reduction of the chiral
blocks results in correlators of rank-two conformal
Jordan cells. In order to argue more directly,
we recall that a conformal Jordan cell of rank $r$ \cite{RAK}
consists of one primary field, $\var_0(z)$, and $r-1$ logarithmic 
and quasi-primary partner fields, $\var_1(z),\dots,\var_{r-1}$,
satisfying
\be
 [L_n,\var_i(z)]\ =\ \left(z^{n+1}\pa_z+\Dh(n+1)z^n\right)
  \var_i(z)+(n+1)z^n\var_{i-1}(z)
\label{r}
\ee
One could say that the field $\var_i(z)$ has degree $i$ or is
at depth $i$. That is, the depth is given by the number of
adjoint actions of the Virasoro modes required to reach the
primary field in the cell. The rank is then given by one plus
the maximum depth. If we extend this characterization to
the affine case, we would say that a field is at depth $i$ if
$i$ adjoint actions of the Lie algebra generators (or more 
generally, $i$ adjoint actions of symmetry generators)
are required to reach a primary field in the affine Jordan cell. 
With the rank denoting one plus the maximum depth just defined,
the rank of our affine Jordan cells is indeed two.

It would be interesting to extend our work to higher ranks in the sense
just indicated. A natural construction seems to suggest itself
and is based on the following simple observation.
The conformal Jordan cell (\ref{r}) may be written compactly
as
\be
 [L_n,\upsilon(z;\thh)]\ =\ \left(z^{n+1}\pa_z+(\Dh+\thh)(n+1)z^n\right)
  \upsilon(z)
\label{upstau}
\ee
where we have introduced the generating-function unified cell
as
\be
 \upsilon(z;\thh)\ =\ \sum_{i=0}^{r-1}\thh^i\var_i(z)
\label{ups}
\ee
In this Section \ref{concl}, $\thh$ is a nilpotent, yet even, parameter satisfying
\be
 \thh^r\ =\ 0,\ \ \ \ \ \ \ \ \ \ \ \ \thh^{r-1}\ \neq\ 0
\label{tauhr}
\ee
We thus suggest to generalize the affine Jordan cell by introducing
the differential-operator realization  
\bea
 D_+(x;\theta)&=&x^2\pa_x-2(j+\theta)x\nn
 D_0(x;\theta)&=&x\pa_x-(j+\theta)\nn
 D_-(x;\theta)&=&-\pa_x
\label{Dtau}
\eea
of the Lie algebra $sl(2)$
and the corresponding generating-function unified cell 
$\upsilon(z,x;\theta)$ satisfying
\be
 [J_a,\upsilon(z,x;\theta)]\ =\ -D_a(x;\theta)\upsilon(z,x;\theta)
\label{JDups}
\ee
In this Section \ref{concl}, 
$\theta$ is a nilpotent, yet even, parameter satisfying
\be
 \theta^r\ =\ 0,\ \ \ \ \ \ \ \ \ \ \ \ \theta^{r-1}\ \neq\ 0
\label{taur}
\ee
The generalization of the expansion (\ref{Upsxexp}) would then read
\be
 \upsilon(z,x;\theta)\ =\ \sum_{i=0}^{r-1}\theta^i\Theta_i(z,x)
\label{upsvar}
\ee
where $\Theta_0(z,x)$ is a generating-function primary field
similar to $\Phi(z,x)$ in (\ref{Upsxexp}).
An examination of hamiltonian reduction of the correlators
based on these higher-rank affine Jordan cells require
knowledge on higher-rank conformal Jordan cells.
Partial results in this direction may be found in \cite{RAK,FloOPE}.
Conformal Jordan cells of infinite rank have been introduced
in \cite{infcell}.

An interesting extension of the work \cite{logWard}
concerns the general solution to the superconformal
Ward identities appearing in logarithmic superconformal
field theory. Results in this direction may be found in 
\cite{KAG}.
A complete solution would facilitate an extension of the
present work to $OSp(1|2)$ WZW models and their
hamiltonian reduction. This deserves to be explored further.

As already mentioned, we hope to address elsewhere 
the classification problem of affine Jordan cells.
In particular, indecomposable representations as
extensions of non-integrable representations would be
interesting to understand. These results could eventually be
extended further to the higher-rank affine Jordan cells
based on (\ref{Dtau}) and (\ref{JDups}) and could be developed along
the lines of Section \ref{affJc}.

We also hope to study the
four-point chiral blocks involving our affine Jordan cells.
It appears straightforward to implement the Ward identities,
after which the general four-point chiral blocks
should follow from the modified KZ equations (\ref{KZth}), (\ref{KZith}).
To test whether the extended prescription for hamiltonian
reduction employed above also applies to these four-point functions,
one could compare the resulting correlators
to the recently obtained results on four-point conformal blocks
\cite{FK,Nag}. 
\vskip.5cm
\noindent{\em Acknowledgements}
\vskip.1cm
\noindent  The author is grateful to M. Flohr for very helpful
comments on the manuscript.

\appendix

\section{Analysis of Ward identities}
\label{app}

Below are indicated some of the steps leading to the general 
expressions for the generating-function
two- and three-point chiral blocks given in (\ref{2xth})
and (\ref{3xth}), respectively.

\subsection{Two-point chiral blocks}

We initially consider the case with two generating-function unified
cells, that is, $\theta_1,\theta_2\neq0$.
Expanding (\ref{2cw}) leads to the conditions
\bea
 0&=&B^0-2(h-\D_1)A^0\ =\ B^0-2(h-\D_2)A^0\nn
 0&=&B^1-2(h-\D_1)A^1+2\frac{2j_1+1}{k+2}A^0
  \ =\ B^1-2(h-\D_2)A^1\nn
 0&=&B^2-2(h-\D_1)A^2
  \ =\ B^2-2(h-\D_2)A^2+2\frac{2j_2+1}{k+2}A^0\nn
 0&=&B^{12}-2(h-\D_1)A^{12}+2\frac{2j_1+1}{k+2}A^2
  \ =\ B^{12}-2(h-\D_2)A^{12}+2\frac{2j_2+1}{k+2}A^1\nn
 0&=&(h-\D_1)B^0\ =\ (h-\D_2)B^0\nn
 0&=&(h-\D_1)B^1-\frac{2j_1+1}{k+2}B^0
  \ =\ (h-\D_2)B^1\nn
 0&=&(h-\D_1)B^2
  \ =\ (h-\D_2)B^2-\frac{2j_2+1}{k+2}B^0\nn
 0&=&(h-\D_1)B^{12}-\frac{2j_1+1}{k+2}B^2
  \ =\ (h-\D_2)B^{12}-\frac{2j_2+1}{k+2}B^1\nn
 0&=&(h-\D_1)C^0\ =\ (h-\D_2)C^0\nn
 0&=&(h-\D_1)C^1-\frac{2j_1+1}{k+2}C^0
  \ =\ (h-\D_2)C^1\nn
 0&=&(h-\D_1)C^2
  \ =\ (h-\D_2)C^2-\frac{2j_2+1}{k+2}C^0\nn
 0&=&(h-\D_1)C^{12}-\frac{2j_1+1}{k+2}C^2
  \ =\ (h-\D_2)C^{12}-\frac{2j_2+1}{k+2}C^1
\label{2cw2}
\eea
whereas an expansion of (\ref{2aw}) yields the conditions
\bea 
 0&=&C^0+2(s-j_1)A^0\ =\ C^0+2(s-j_2)A^0\nn
 0&=&C^1+2(s-j_1)A^1-2A^0\ =\ C^1+2(s-j_2)A^1\nn
 0&=&C^2+2(s-j_1)A^2\ =\ C^2+2(s-j_2)A^2-2A^0\nn
 0&=&C^{12}+2(s-j_1)A^{12}-2A^2\ =\ C^{12}+2(s-j_2)A^{12}-2A^1\nn
 0&=&(s-j_1)B^0\ =\ (s-j_2)B^0\nn
 0&=&(s-j_1)B^1-B^0\ =\ (s-j_2)B^1\nn
 0&=&(s-j_1)B^2\ =\ (s-j_2)B^2-B^0\nn
 0&=&(s-j_1)B^{12}-B^2\ =\ (s-j_2)B^{12}-B^1\nn
 0&=&(s-j_1)C^0\ =\ (s-j_2)C^0\nn
 0&=&(s-j_1)C^1-C^0\ =\ (s-j_2)C^1\nn
 0&=&(s-j_1)C^2\ =\ (s-j_2)C^2-C^0\nn
 0&=&(s-j_1)C^{12}-C^2\ =\ (s-j_2)C^{12}-C^1
\label{2aw2}
\eea
It follows immediately that a non-trivial solution requires
\be
 s\ =\ j_1\ =\ j_2,\ \ \ \ \ \ \ h\ =\ \D_1\ =\ \D_2
\label{sh}
\ee
further implying the relations
\bea
 0&=&A^0\ =\ B^0\ =\ B^1\ =\ B^2\ =\ C^0\ =\ C^1\ =\ C^2\nn
 0&=&A^1-A^2\ =\ B^{12}+2\frac{2j_1+1}{k+2}A^1\ =\ C^{12}-2A^1
\label{2cond2}
\eea
The parameter $A^{12}$ is independent of the other ones.

In the case where $\theta_1\neq0$ while $\theta_2=0$,
the third conformal Ward identity (i.e., (\ref{2cw})) yields
\bea
 0&=&B^0-2(h-\D_1)A^0\ =\ B^0-2(h-\D_2)A^0\nn
 0&=&B^1-2(h-\D_1)A^1+2\frac{2j_1+1}{k+2}A^0
  \ =\ B^1-2(h-\D_2)A^1\nn
 0&=&(h-\D_1)B^0\ =\ (h-\D_2)B^0\nn
 0&=&(h-\D_1)B^1-\frac{2j_1+1}{k+2}B^0
  \ =\ (h-\D_2)B^1\nn
 0&=&(h-\D_1)C^0\ =\ (h-\D_2)C^0\nn
 0&=&(h-\D_1)C^1-\frac{2j_1+1}{k+2}C^0
  \ =\ (h-\D_2)C^1
\label{2cw1}
\eea
while the third $SL(2,\mathbb{R})$ Ward identity (i.e., (\ref{2aw}))
corresponds to
\bea 
 0&=&C^0+2(s-j_1)A^0\ =\ C^0+2(s-j_2)A^0\nn
 0&=&C^1+2(s-j_1)A^1-2A^0\ =\ C^1+2(s-j_2)A^1\nn
 0&=&(s-j_1)B^0\ =\ (s-j_2)B^0\nn
 0&=&(s-j_1)B^1-B^0\ =\ (s-j_2)B^1\nn
 0&=&(s-j_1)C^0\ =\ (s-j_2)C^0\nn
 0&=&(s-j_1)C^1-C^0\ =\ (s-j_2)C^1
\label{2aw1}
\eea
It is stressed that $B^2$, for example, does not exist
(or is set to zero) in this case and should therefore not
be treated as a free parameter. As above, the spins and
weights are seen to satisfy (\ref{sh}), and it follows that
\be
 0\ =\ A^0\ =\ B^0\ =\ B^1\ =\ C^0\ =\ C^1
\label{2cond1}
\ee
while $A^1$ is the only free parameter.

In the case where $\theta_1=\theta_2=0$,
the two sets of conditions reduce to
\bea
 0&=&B^0-2(h-\D_1)A^0\ =\ B^0-2(h-\D_2)A^0\nn
 0&=&(h-\D_1)B^0\ =\ (h-\D_2)B^0\nn
 0&=&(h-\D_1)C^0\ =\ (h-\D_2)C^0
\label{2cw0}
\eea
and
\bea 
 0&=&C^0+2(s-j_1)A^0\ =\ C^0+2(s-j_2)A^0\nn
 0&=&(s-j_1)B^0\ =\ (s-j_2)B^0\nn
 0&=&(s-j_1)C^0\ =\ (s-j_2)C^0
\label{2aw0}
\eea
Once again, the spins and weights satisfy (\ref{sh}).
This time, $B^0=C^0=0$ while $A^0$ is 
the only free parameter.

This analysis leads to the two-point chiral blocks given in (\ref{2x}).

\subsection{Three-point chiral blocks}

Based on the ansatz (\ref{3ansx}), the third conformal Ward
identity (\ref{confwardx}) corresponds to the conditions
\bea
 0&=&(s_1+s_3-2j_1-2\theta_1)A+C_{12}+C_{13}
  \ =\ (s_1+s_2-2j_2-2\theta_2)A+C_{12}+C_{23}\nn
 &=&(s_2+s_3-2j_3-2\theta_3)A+C_{23}+C_{13}\nn
 0&=&(s_1+s_3-2j_1-2\theta_1)B_{12}+E_{11}+E_{13}
  \ =\ (s_1+s_2-2j_2-2\theta_2)B_{12}+E_{11}+E_{12}\nn
 &=&(s_2+s_3-2j_3-2\theta_3)B_{12}+E_{12}+E_{13}\nn
 0&=&(s_1+s_3-2j_1-2\theta_1)B_{23}+E_{21}+E_{23}
  \ =\ (s_1+s_2-2j_2-2\theta_2)B_{23}+E_{21}+E_{22}\nn
 &=&(s_2+s_3-2j_3-2\theta_3)B_{23}+E_{22}+E_{23}\nn
 0&=&(s_1+s_3-2j_1-2\theta_1)B_{13}+E_{31}+E_{33}
  \ =\ (s_1+s_2-2j_2-2\theta_2)B_{13}+E_{31}+E_{32}\nn
 &=&(s_2+s_3-2j_3-2\theta_3)B_{13}+E_{32}+E_{33}\nn
 0&=&(s_1+s_3-2j_1-2\theta_1)C_{12}+2F_{11}+F_{13}
  \ =\ (s_1+s_2-2j_2-2\theta_2)C_{12}+2F_{11}+F_{12}\nn
 &=&(s_2+s_3-2j_3-2\theta_3)C_{12}+F_{12}+F_{13}\nn
 0&=&(s_1+s_3-2j_1-2\theta_1)C_{23}+F_{12}+F_{23}
  \ =\ (s_1+s_2-2j_2-2\theta_2)C_{23}+F_{12}+2F_{22}\nn
 &=&(s_2+s_3-2j_3-2\theta_3)C_{23}+2F_{22}+F_{23}\nn
 0&=&(s_1+s_3-2j_1-2\theta_1)C_{13}+F_{13}+2F_{33}
  \ =\ (s_1+s_2-2j_2-2\theta_2)C_{13}+F_{13}+F_{23}\nn
 &=&(s_2+s_3-2j_3-2\theta_3)C_{13}+F_{23}+2F_{33}\nn
 0&=&(s_1+s_3-2j_1-2\theta_1)D_{ij}
  \ =\ (s_1+s_2-2j_2-2\theta_2)D_{ij}
  \ =\ (s_2+s_3-2j_3-2\theta_3)D_{ij}\nn
0&=&(s_1+s_3-2j_1-2\theta_1)E_{ij}
  \ =\ (s_1+s_2-2j_2-2\theta_2)E_{ij}
  \ =\ (s_2+s_3-2j_3-2\theta_3)E_{ij}\nn
0&=&(s_1+s_3-2j_1-2\theta_1)F_{ij}
  \ =\ (s_1+s_2-2j_2-2\theta_2)F_{ij}
  \ =\ (s_2+s_3-2j_3-2\theta_3)F_{ij}
\label{3cw}
\eea
whereas the third $SL(2,\mathbb{R})$ Ward identity corresponds to the conditions
\bea
 0&=&(-h_1-h_3+2\D_1+2\mu_1)A+B_{12}+B_{13}
  \ =\ (-h_1-h_2+2\D_2+2\mu_2)A+B_{12}+B_{23}\nn
  &=&(-h_2-h_3+2\D_3+2\mu_3)A+B_{23}+B_{13}\nn
 0&=&(-h_1-h_3+2\D_1+2\mu_1)B_{12}+2D_{11}+D_{13}
  \ =\ (-h_1-h_2+2\D_2+2\mu_2)B_{12}+2D_{11}+D_{12}\nn
  &=&(-h_2-h_3+2\D_3+2\mu_3)B_{12}+D_{12}+D_{13}\nn
 0&=&(-h_1-h_3+2\D_1+2\mu_1)B_{23}+D_{12}+D_{23}
  \ =\ (-h_1-h_2+2\D_2+2\mu_2)B_{23}+D_{12}+2D_{22}\nn
  &=&(-h_2-h_3+2\D_3+2\mu_3)B_{23}+2D_{22}+D_{23}\nn
 0&=&(-h_1-h_3+2\D_1+2\mu_1)B_{13}+D_{13}+2D_{33}
  \ =\ (-h_1-h_2+2\D_2+2\mu_2)B_{13}+D_{13}+D_{23}\nn
  &=&(-h_2-h_3+2\D_3+2\mu_3)B_{13}+D_{23}+2D_{33}\nn
 0&=&(-h_1-h_3+2\D_1+2\mu_1)C_{12}+E_{11}+E_{31}
  \ =\ (-h_1-h_2+2\D_2+2\mu_2)C_{12}+E_{11}+E_{21}\nn
  &=&(-h_2-h_3+2\D_3+2\mu_3)C_{12}+E_{21}+E_{31}\nn
 0&=&(-h_1-h_3+2\D_1+2\mu_1)C_{23}+E_{12}+E_{32}
  \ =\ (-h_1-h_2+2\D_2+2\mu_2)C_{23}+E_{12}+E_{22}\nn
  &=&(-h_2-h_3+2\D_3+2\mu_3)C_{23}+E_{22}+E_{32}\nn
 0&=&(-h_1-h_3+2\D_1+2\mu_1)C_{13}+E_{13}+E_{33}
  \ =\ (-h_1-h_2+2\D_2+2\mu_2)C_{13}+E_{13}+E_{23}\nn
  &=&(-h_2-h_3+2\D_3+2\mu_3)C_{13}+E_{23}+E_{33}\nn
 0&=&(-h_1-h_3+2\D_1+2\mu_1)D_{ij}
  \ =\ (-h_1-h_2+2\D_2+2\mu_2)D_{ij}
  \ =\ (-h_2-h_3+2\D_3+2\mu_3)D_{ij}\nn
 0&=&(-h_1-h_3+2\D_1+2\mu_1)E_{ij}
  \ =\ (-h_1-h_2+2\D_2+2\mu_2)E_{ij}
  \ =\ (-h_2-h_3+2\D_3+2\mu_3)E_{ij}\nn
 0&=&(-h_1-h_3+2\D_1+2\mu_1)F_{ij}
  \ =\ (-h_1-h_2+2\D_2+2\mu_2)F_{ij}
  \ =\ (-h_2-h_3+2\D_3+2\mu_3)F_{ij}\nn
\label{3aw} 
\eea
To keep the notation simple, we have left out the explicit indications
that the structure constants depend on the $\theta$s.
To keep the presentation simple as well, we will leave out most of the
details in the analysis of these conditions. Since the approach
essentially is the same
as the one employed in the study of two-point chiral blocks, we will merely
outline the main steps.

We distinguish between the different numbers of unified
cells, that is, the different numbers of non-vanishing $\theta$s.
In every case, one finds the relations
\bea
 s_1\ =\ j_1+j_2-j_3,\ \ \ \ \ \ \ s_2&=&-j_1+j_2+j_3,\ \ \ \ \ \ \ 
  s_3\ =\ j_1-j_2+j_3\nn
 h_1\ =\ \D_1+\D_2-\D_3,\ \ \ \ \ \ \ h_2&=&-\D_1+\D_2+\D_3,
  \ \ \ \ \ \ \ \ h_3\ =\ \D_1-\D_2+\D_3
\label{sjhD}
\eea
Having split the analysis into the four cases characterized
by 0, 1, 2 or 3 unified cells, one expands the conditions
(\ref{3cw}) and (\ref{3aw}) on the set of associated $\theta$s,
where it is recalled that $\mu_i=(2j_i+1)\theta_i/(k+2)$.
Since the resulting conditions are linear in the structure
constants $A^0,B_{ij}^0$ etc, it is straightforward to work out
the relations between the structure constants associated
to the four cases. The relations are listed below.

In the case where $\theta_1=\theta_2=\theta_3=0$, we find
\be
 B_{ij}^0\ =\ C_{ij}^0\ =\ D_{ij}^0\ =\ E_{ij}^0\ =\ F_{ij}^0\ =\ 0
\ee 
which means that $A^0$ is the only free structure constant.

In the case where $\theta_2=\theta_3=0$ while $\theta_1\neq0$, we find
\bea
 &&0\ =\ B_{ij}^0\ =\ C_{ij}^0\ =\ D_{ij}^{0}\ =\ D_{ij}^1
 \ =\ E_{ij}^{0}\ =\ E_{ij}^1\ =\ F_{ij}^{0}\ =\ F_{ij}^1\nn
 &&B_{12}^1\ =\ -B_{23}^1\ =\ B_{13}^1\ =\ -\frac{2j_1+1}{k+2}A^0\nn
 &&C_{12}^1\ =\ -C_{23}^1\ =\ C_{13}^1\ =\ A^0
\label{31x}
\eea 
while $A^1$ is unconstrained. That is, we may consider $A^0$ and $A^1$ as
the only independent structure constants.

In the case where $\theta_3=0$ while $\theta_1,\theta_2\neq0$, we find
\bea
 &&0\ =\ B_{ij}^0\ =\ C_{ij}^0\ =\ D_{ij}^0\ =\ D_{ij}^1\ =\ D_{ij}^2\ =\ E_{ij}^0
  \ =\ E_{ij}^1\ =\ E_{ij}^2\ =\ F_{ij}^0\ =\ F_{ij}^1\ =\ F_{ij}^2\nn
 &&B_{12}^1\ =\ -B_{23}^1\ =\ B_{13}^1\ =\ -\frac{2j_1+1}{k+2}A^0,
 \ \ \ \ \ \ \ B_{12}^2\ =\ B_{23}^1\ =\ -B_{13}^1\ =\ -\frac{2j_2+1}{k+2}A^0\nn
 &&B_{12}^{12}\ =\ -\frac{2j_2+1}{k+2}A^1-\frac{2j_1+1}{k+2}A^2,
  \ \ \ \ \ \ \ B_{23}^{12}\ =\ -B_{13}^{12}\ =\ 
   -\frac{2j_2+1}{k+2}A^1+\frac{2j_1+1}{k+2}A^2\nn
 &&C_{12}^1\ =\ -C_{23}^1\ =\ C_{13}^1\ =\ A^0,\ \ \ \ \ \ \ 
  C_{12}^2\ =\ C_{23}^2\ =\ -C_{13}^2\ =\ A^0\nn
 &&C_{12}^{12}\ =\ A^1+A^2,\ \ \ \ \ \ \ C_{23}^{12}\ =\ -C_{13}^{12}
  \ =\ A^1-A^2\nn
 &&D_{11}^{12}\ =\ -D_{22}^{12}\ =\ \hf D_{23}^{12}\ =\ -D_{33}^{12}
  \ =\ \frac{(2j_1+1)(2j_2+1)}{(k+2)^2}A^0,\ \ \ \ \ \ \ 
  D_{12}^{12}\ =\ D_{13}^{12}\ =\ 0\nn
 &&E_{11}^{12}\ =\ -E_{22}^{12}\ =\ E_{23}^{12}\ =\ E_{32}^{12}
  \ =\ -E_{33}^{12}\ =\ -2\frac{j_1+j_2+1}{k+2}A^0\nn
 &&E_{12}^{12}\ =\ -E_{13}^{12}\ =\ -E_{21}^{12}\ =\ E_{31}^{12}
  \ =\ \frac{-2j_1+2j_2}{k+2}A^0\nn
 &&F_{11}^{12}\ =\ -F_{22}^{12}\ =\ \hf F_{23}^{12}\ =\ -F_{33}^{12}
  \ =\ A^0,\ \ \ \ \ \ \ F_{12}^{12}\ =\ F_{13}^{12}\ =\ 0
\label{32x}
\eea
while $A^{12}$ is unconstrained. That is, we may consider $A^0$,
$A^1$, $A^2$ and $A^{12}$ as the only independent structure constants.

In the case where $\theta_1,\theta_2,\theta_3\neq0$, we find
\bea
 &&0\ =\ A^0\ =\ B_{ij}^0\ =\ B_{ij}^{l}\ =\ C_{ij}^0\ =\ C_{ij}^l\nn
 &&0\ =\ D_{ij}^0\ =\ D_{ij}^l\ =\ D_{ij}^{lm}\ =\ E_{ij}^0\ =\ E_{ij}^l\ =\ E_{ij}^{lm}
   \ =\ F_{ij}^0\ =\ F_{ij}^l\ =\ F_{ij}^{lm}\nn
 &&B_{12}^{12}\ =\ -\frac{2j_2+1}{k+2}A^1-\frac{2j_1+1}{k+2}A^2,\ \ \ \ \ \ \ 
  B_{23}^{12}\ =\ -B_{13}^{12}
   \ =\ -\frac{2j_2+1}{k+2}A^1+\frac{2j_1+1}{k+2}A^2\nn
 &&B_{23}^{23}\ =\ -\frac{2j_3+1}{k+2}A^2-\frac{2j_2+1}{k+2}A^3,\ \ \ \ \ \ \ 
  B_{12}^{23}\ =\ -B_{13}^{23}
   \ =\ \frac{2j_3+1}{k+2}A^2-\frac{2j_2+1}{k+2}A^3\nn
 &&B_{13}^{13}\ =\ -\frac{2j_3+1}{k+2}A^1-\frac{2j_1+1}{k+2}A^3,\ \ \ \ \ \ \ 
  B_{12}^{13}\ =\ -B_{23}^{13}
   \ =\ \frac{2j_3+1}{k+2}A^1-\frac{2j_1+1}{k+2}A^3\nn
 &&B_{12}^{123}\ =\ \frac{2j_3+1}{k+2}A^{12}-\frac{2j_1+1}{k+2}A^{23}
   -\frac{2j_2+1}{k+2}A^{13}\nn
 &&B_{23}^{123}\ =\ -\frac{2j_3+1}{k+2}A^{12}+\frac{2j_1+1}{k+2}A^{23}
   -\frac{2j_2+1}{k+2}A^{13}\nn
 &&B_{12}^{123}\ =\ -\frac{2j_3+1}{k+2}A^{12}-\frac{2j_1+1}{k+2}A^{23}
   +\frac{2j_2+1}{k+2}A^{13}\nn
 &&C_{12}^{12}\ =\ A^1+A^2,\ \ \ \ \ \ \ C_{23}^{12}\ =\ -C_{13}^{12}\ =\ A^1-A^2\nn
 &&C_{23}^{23}\ =\ A^2+A^3,\ \ \ \ \ \ \ C_{12}^{23}\ =\ -C_{13}^{23}\ =\ -A^2+A^3\nn
 &&C_{13}^{13}\ =\ A^1+A^3,\ \ \ \ \ \ \ C_{12}^{13}\ =\ -C_{23}^{13}\ =\ -A^1+A^3\nn
 &&C_{12}^{123}\ =\ -A^{12}+A^{23}+A^{13},\ \ \ \ \ \ \ 
  C_{23}^{123}\ =\ A^{12}-A^{23}+A^{13},\ \ \ \ \ \ \ 
  C_{13}^{123}\ =\ A^{12}+A^{23}-A^{13}\nn 
 &&D_{11}^{123}\ =\ -\frac{(2j_2+1)(2j_3+1)}{(k+2)^2}A^1
   -\frac{(2j_1+1)(2j_3+1)}{(k+2)^2}A^2+\frac{(2j_1+1)(2j_2+1)}{(k+2)^2}A^3\nn
 &&D_{22}^{123}\ =\ \frac{(2j_2+1)(2j_3+1)}{(k+2)^2}A^1
   -\frac{(2j_1+1)(2j_3+1)}{(k+2)^2}A^2-\frac{(2j_1+1)(2j_2+1)}{(k+2)^2}A^3\nn
 &&D_{33}^{123}\ =\ -\frac{(2j_2+1)(2j_3+1)}{(k+2)^2}A^1
   +\frac{(2j_1+1)(2j_3+1)}{(k+2)^2}A^2-\frac{(2j_1+1)(2j_2+1)}{(k+2)^2}A^3\nn
 &&D_{12}^{123}\ =\ 2\frac{(2j_1+1)(2j_3+1)}{(k+2)^2}A^2,\ \ \ \ \ \ \ 
  D_{23}^{123}\ =\ 2\frac{(2j_1+1)(2j_2+1)}{(k+2)^2}A^3\nn 
 &&D_{13}^{123}\ =\ 2\frac{(2j_2+1)(2j_3+1)}{(k+2)^2}A^1\nn
 &&E_{11}^{123}\ =\ \frac{2(j_2+j_3+1)}{k+2}A^1
   +\frac{2(j_1+j_3+1)}{k+2}A^2-\frac{2(j_1+j_2+1)}{k+2}A^3\nn
 &&E_{12}^{123}\ =\ \frac{-2j_2+2j_3}{k+2}A^1
   -\frac{2(j_1+j_3+1)}{k+2}A^2+\frac{-2j_1+2j_2}{k+2}A^3\nn
&&E_{13}^{123}\ =\ -\frac{2(j_2+j_3+1)}{k+2}A^1
   +\frac{-2j_1+2j_3}{k+2}A^2+\frac{2j_1-2j_2}{k+2}A^3\nn
&&E_{21}^{123}\ =\ \frac{2j_2-2j_3}{k+2}A^1
   -\frac{2(j_1+j_3+1)}{k+2}A^2+\frac{2j_1-2j_2}{k+2}A^3\nn
&&E_{22}^{123}\ =\ -\frac{2(j_2+j_3+1)}{k+2}A^1
   +\frac{2(j_1+j_3+1)}{k+2}A^2+\frac{2(j_1+j_2+1)}{k+2}A^3\nn
&&E_{23}^{123}\ =\ \frac{-2j_2+2j_3}{k+2}A^1
   +\frac{2j_1-2j_3}{k+2}A^2-\frac{2(j_1+j_2+1)}{k+2}A^3\nn
&&E_{31}^{123}\ =\ -\frac{2(j_2+j_3+1)}{k+2}A^1
   +\frac{2j_1-2j_3}{k+2}A^2+\frac{-2j_1+2j_2}{k+2}A^3\nn
&&E_{32}^{123}\ =\ \frac{2j_2-2j_3}{k+2}A^1
   +\frac{-2j_1+2j_3}{k+2}A^2-\frac{2(j_1+j_2+1)}{k+2}A^3\nn
&&E_{33}^{123}\ =\ \frac{2(j_2+j_3+1)}{k+2}A^1
   -\frac{2(j_1+j_3+1)}{k+2}A^2+\frac{2(j_1+j_2+1)}{k+2}A^3\nn
&&F_{11}^{123}\ =\ -A^1-A^2+A^3,\ \ \ \ \ \ \ 
  F_{22}^{123}\ =\ A^1-A^2-A^3,\ \ \ \ \ \ \ 
  F_{33}^{123}\ =\ -A^1+A^2-A^3\nn
&&F_{12}^{123}\ =\ 2A^2,\ \ \ \ \ \ \ 
 F_{23}^{123}\ =\ 2A^3,\ \ \ \ \ \ \ F_{13}^{123}\ =\ 2A^1
\label{33x}
\eea
while $A^{123}$ is unconstrained. That is, we may consider $A^0$,
$A^i$, $A^{ij}$ and $A^{123}$ as the only independent structure constants.

The relations corresponding to the situation where
$\theta_1=\theta_3=0$ while $\theta_2\neq0$, for example,
are obtained from the relations corresponding to the case
where $\theta_2=\theta_3=0$ while $\theta_1\neq0$ by an 
appropriate permutation in the indices.

These results lead to the three-point chiral blocks given
in (\ref{3xth}).

\end{document}